\documentclass[10pt,twocolumn,twoside]{IEEEtran}
\usepackage{amsmath,graphicx}
\usepackage{amssymb}
\usepackage{graphicx}
\usepackage{graphics}
\usepackage{epstopdf}
\usepackage{amsmath}
\usepackage{algorithmicx}
\usepackage{algorithm}
\usepackage{algpseudocode}
\usepackage{multirow}
\usepackage{bm}
\usepackage{enumerate}
\usepackage{paralist}
\usepackage{array}
\begin{document}

\title{Sensor scheduling with time, energy and communication constraints}

\author{Cristian Rusu, John Thompson and Neil M. Robertson \thanks{C. Rusu is with the School of Computing, the National College of Ireland, Ireland. J. Thompson is with the Institute for Digital Communications, School of Engineering, The University of Edinburgh, UK. N. M. Robertson is with Queen's University of Belfast, UK. Emails: cristian.rusu@ncirl.ie, john.thompson@ed.ac.uk, n.robertson@qub.ac.uk. Demo source code available online http://udrc.eng.ed.ac.uk/sites/udrc.eng.ed.ac.uk/files/attachments/demo\_sensor \_management.zip
		
		The authors acknowledge support from the Engineering and Physical Sciences Research Council (EPSRC) [EP/K014277/1] and the MOD University Defence Research Collaboration (UDRC) on Signal Processing.}}

\maketitle

\begin{abstract}
In this paper we present new algorithms and analysis for the linear inverse sensor placement and scheduling problems over multiple time instances with power and communications constraints. The proposed algorithms, which deal directly with minimizing the mean squared error (MSE), are based on the convex relaxation approach to address the binary optimization scheduling problems that are formulated in sensor network scenarios. We propose to balance the energy and communications demands of operating a network of sensors over time while we still guarantee a minimum level of estimation accuracy. We measure this accuracy by the MSE for which we provide average case and lower bounds analyses that hold in general, irrespective of the scheduling algorithm used. We show experimentally how the proposed algorithms perform against state-of-the-art methods previously described in the literature.
\end{abstract}


\section{Introduction}

Sensor networks are often used to measure and monitor physical phenomena like temperature, humidity and concentration of certain pollutants in an area of interest over time \cite{Leus2016}. Modern wireless sensor networks may be composed of a large number of heterogeneous sensors each with its own (possibly limited) power supply capable of performing measurements, processing the result and communicating it to other neighboring sensors in the network at regular times. In this paper we consider the situation where, without any particular assumptions on the parameters to be estimated, the measurements taken by the sensor network are used to solve a linear inverse problem. In this setting, the problem of selecting only a subset of the available sensors while ensuring a certain level of estimation accuracy has been extensively studied in the literature.

Sensor selection (or sensor placement) \cite{Hero2, Hero1} is of central importance when considering the classical problem of parameter estimation from a given set of linear measurements that describe an operational sensor network. Given a fixed set of potential locations, the sensor placement problem asks where the sensors should be placed in order to maximize on average the estimation accuracy of the network. Once the most informative locations are identified the sensors are placed in their locations for the whole lifetime of the network. If we now consider a network where each sensor has a particular energy and communication profile and is capable of performing a measurement with a particular quality, an interesting problem that arises is how to schedule each sensor over time such that the estimation accuracy of the network is never worse than a prescribed level while we also control the energy consumption. This is done for example to make sure that no sensor goes off-line due to overuse. We call this the sensor scheduling problem. In this paper we tackle both problems and propose new heuristics to address them and provide new theoretical insights into their behavior. Because the problems are combinatorial in nature (they amount to mixed-integer optimization problems) they are NP-hard to solve exactly in general. Therefore, following previous literature, we settle on proposing sub-optimal but numerically efficient algorithms and comparing them with previously proposed methods.

The sensor placement (and in general the sensor management) problems have been extensively studied in the past. A general approach is to use greedy methods based on a minimum eigenspace approach \cite{MPME2016} or with submodularity based performance guarantees \cite{Vikalo2010} that provide results within $(1-e^{-1})$ of the optimal solution. Another popular greedy sensor selection method, called FrameSense \cite{FrameSense2014}, initially activates all the sensors and then removes one at each step based on a ``worst-out" principle to optimize its submodular objective function. Convex optimization techniques have also been proven useful for experimental design \cite[Chapter~7.5]{CO} with $\ell_1$ \cite{JoshiBoyd2009, Varshney2014b} and reweighted $\ell_1$ norm minimization \cite{IRL1} approaches such as \cite{ChepuriLeusVeen2013, LiuVempatyFardadMasazadeVarshney2014, SparSenSe2014, ChepuriLeus2015}. Earlier work used information theoretic approaches like mutual information maximization \cite{WangEstrin2004, Guestrin2005} and cross entropy optimization \cite{Naeem2009} or other search heuristics like genetic algorithms \cite{Search1}, tabu search \cite{Search2} and branch-and-bound methods \cite{BandB} to solve the sensor placement problems. Several recent works have also considered non-linear sensor networks \cite{SparSenSe2014}, tracking applications \cite{Masazade2012, NonLinearFiltering}, distributed sensing scenarios \cite{SparseAware, DistributedSensors}, correlated noise models \cite{CorrelatedNoise}, estimation of continuous variables \cite{Continuous}. Additional scenarios where further limitations are added to the network sensing problem include: energy budget constraints \cite{Mo2011} and ways to maximize the lifetime per unit cost in wireless sensor networks \cite{ChenChuahZhao2005}, $\ell_2$ regularization terms that discourage the selection of the same sensors over a period of time \cite{LiuVempatyFardadMasazadeVarshney2014, LiuChenVempatyFardadShenVarshney2015} and scheduling \cite{Scheduling2012, Scheduling2016} over the network. In the same spirit, recent work introduces a greedy sampling set selection algorithm for graph signal processing applications \cite{GreedySetSelectionGraph2016}.

The contribution of this paper is two-fold.

The first contribution is to propose an $\ell_\infty$ regularized iteratively reweighted $\ell_1$ sensor scheduling algorithm \cite{BalancedSensorManagement}. The approach is based on the previously introduced $\ell_1$ convex norm optimization approach. We are able to accommodate energy and communications constraints in order to balance the estimation accuracy of the sensor network over multiple time instances with its energy consumption. Solving the proposed convex optimization process is numerically efficient since it can be done in polynomial time by off-the-shelf solvers \cite{CVX}. This approach is slower than some previously proposed greedy methods but it has the advantage of easily accommodating real-world constraints of the sensor networks.

The second contribution is to provide a framework where an average analysis of the placement problem and a worst case analysis for the scheduling problem are performed, both in the case where the overall sensor network is described by a tight frame or measurement matrix. This general analysis provides insights into the empirical possible performance of sensor scheduling \textit{independent of the algorithms used}. We also provide connections to other related research areas where bounds on the eigenvalues of sub-matrices of a matrix have been developed. The case of a tight measurement matrix, although not often seen in practical sensor networks problems, is of interest since it is the optimal choice in linear inverse problems and also appears in sampling set selection for bandlimited graph signals \cite{EfficientSamplingSetSelection2016}, both under noise.

Section II describes the measurement setup we consider; Section III presents the previously proposed methods in the literature for the sensor placement problem; Section IV proposes a new algorithms for the sensor placement and scheduling problems; Section V provides a theoretical analysis of the average estimation accuracy for the sensor placement problem and Section VI shows the numerical results where we compare the proposed method with the state-of-the-art methods from the literature.

\section{The problem setup}

Let us assume that we want to estimate a parameter vector $\mathbf{x}_t$ of size $n$ that changes over $t = 1,\dots,T,$ time instances from $n \leq k_t \leq m$ linear measurements that are given by
\begin{equation}
	\mathbf{y}_t = \mathbf{A}_t\mathbf{x}_t + \mathbf{n}_t,
\end{equation}
where the noise vector $\mathbf{n}_t$ of size $k_t$ is a zero-mean i.i.d. Gaussian vector with variance $\sigma^2 \mathbf{I}$. The rows of the measurement matrix $\mathbf{A}_t$ at time $t$ are chosen from the rows of an overall full rank measurement matrix $\mathbf{A}$ of size $m \times n$. The $i^\text{th}$ row of $\mathbf{A}$ corresponds to the linear measurement performed by the $i^\text{th}$ sensor of the network. Therefore, the matrix $\mathbf{A}$ characterizes the full sensor network which is made up of $m$ elements and the measurement matrices $\mathbf{A}_t$ are subsets of the rows from $\mathbf{A}$. Assuming $k_t \geq n$ sensors are used at time $t$, the least squares estimate are given by
\begin{equation}
	\mathbf{\hat{x}}_t = \mathbf{A}_t^\dagger \mathbf{y}_t = (\mathbf{A}_t^T \mathbf{A}_t)^{-1} \mathbf{A}_t^T \mathbf{y}_t.
	\label{eq:lsrecovery}
\end{equation}

With the understanding that with increased noise levels the estimation accuracy decreases on average, in order to simplify the exposition of the results herein we only consider the fixed noise level $\sigma^2 = 1$. We focus on a full rank measurement matrix $\mathbf{A}_t \in \mathbb{R}^{k_t \times n}$, that represents the $k_t$ sensors that perform linear measurements at time $t$, for which there are several ways to quantify its recovery performance in \eqref{eq:lsrecovery}:
\begin{enumerate}
	\item A-optimality: mean squared error (MSE)
		\begin{equation}
			\text{MSE}(\mathbf{A}_t) =  \text{tr}((\mathbf{A}_t^T \mathbf{A}_t)^{-1})  = \sum_{i=1}^n \frac{1}{\lambda_i(\mathbf{A}_t^T\mathbf{A}_t)}.
			\label{eq:mse}
		\end{equation}
		
	\item E-optimality: worst case error variance (WCE)
		\begin{equation}
			\text{WCE}(\mathbf{A}_t) = \lambda_1 ((\mathbf{A}_t^T \mathbf{A}_t)^{-1})  = \frac{1}{\lambda_n(\mathbf{A}_t^T\mathbf{A}_t)}.
			\label{eq:wce}
		\end{equation}
		
	\item D-optimality: volume of the confidence ellipsoid (VCE)
\end{enumerate}
		\begin{equation}
			\! \! \text{VCE}(\mathbf{A}_t) \! = \! \log \! \det(\mathbf{A}_t^T \mathbf{A}_t) \! = \! \log \! \left( \prod_{i=1}^n \lambda_i(\mathbf{A}_t^T\mathbf{A}_t) \right). \! \! \! \!
			\label{eq:vce}
		\end{equation}
We have denoted here $\lambda_i(\mathbf{A}_t^T\mathbf{A}_t)$ as the $i^\text{th}$ eigenvalue of the symmetric positive-semidefinite matrix $\mathbf{A}_t^T \mathbf{A}_t$ and we assume without loss of generality the ordering $\lambda_1(\mathbf{A}_t^T\mathbf{A}_t) \geq \dots \geq \lambda_n(\mathbf{A}_t^T\mathbf{A}_t) \geq 0$.

Notice that these performance indicators are related as we have $\text{MSE}(\mathbf{A}_t) \leq n \text{WCE}(\mathbf{A}_t)$, while maximizing the $\text{VCE}(\mathbf{A}_t)$ we also maximize the denominator of the $\text{MSE}(\mathbf{A}_t)$ in \eqref{eq:mse} -- but we do not also control the numerator term in \eqref{eq:mse}. Among all the measurement matrices of the same size $k_t \times n$, and for the same Frobenius norm, these performance measures are optimized for $\alpha-$tight frames. A finite frame $\mathbf{A}_t \in \mathbb{R}^{k_t \times n}$ is a countable collection of $k_t$ vectors of size $n$ such that
\begin{equation}
	\alpha \| \mathbf{x} \|_2^2 \leq  \sum_{i=1}^{k_t} | \mathbf{a}_i^T \mathbf{x} |^2 \leq \beta \| \mathbf{x} \|_2^2, \ \forall \ \mathbf{x} \in \mathbb{R}^n,
\end{equation}
where $0 < \alpha \leq \beta < \infty$ are the frame bounds. When $\alpha = \beta$ the frame $\mathbf{A}_t$ is called $\alpha-$tight, or just tight. Tight measurement matrices $\mathbf{A}_t$ obey $\mathbf{A}_t^T\mathbf{A}_t = \alpha \mathbf{I}$ and therefore we have that
\begin{equation}
	\begin{aligned}
	&\text{MSE}(\textbf{A}_t) = \frac{n}{\alpha} = \frac{n^2}{\| \mathbf{A}_t \|_F^2},\
	\text{WCE}(\textbf{A}_t) = \frac{1}{\alpha} = \frac{n}{\| \mathbf{A}_t \|_F^2}, \\
	& \quad \quad \text{VCE}(\mathbf{A}_t) = \log \left( \alpha^n \right) = n \log \left(\frac{\| \mathbf{A}_t \|_F^2}{n} \right),
	\end{aligned}
	\label{eq:tightperformance}
\end{equation}
since $\| \mathbf{A}_t \|_F^2 = \text{tr}(\mathbf{A}_t^T \mathbf{A}_t) = n \alpha$. Higher Frobenius norm (and therefore also higher $\alpha$) of the measurement matrix is equivalent to increasing on average the SNR of the measurements and therefore should lead to better recovery performance in general -- for example $\text{MSE}(\delta \mathbf{A}_t) = \delta^{-2}\text{MSE}(\mathbf{A}_t), \ \delta \in \mathbb{R}$. Since tight frames are optimum for these criteria then for all non-tight frames these values are lower bounds.

To achieve low error indicators, in terms of the eigenvalues of $\mathbf{A}_t^T \mathbf{A}_t$ our goal is twofold:
\begin{itemize}
	\item first, increase the smallest eigenvalue
	\begin{equation}
		\lambda_n(\mathbf{A}_t^T \mathbf{A}_t) \gg 0,
		\label{eq:invertible}
	\end{equation}
	
	\item second, group all eigenvalues such that
	\begin{equation}
	\lambda_i(\mathbf{A}_t^T \mathbf{A}_t) \approx \lambda_j(\mathbf{A}_t^T \mathbf{A}_t),\ \forall i \neq j,
		\label{eq:almosttight}
	\end{equation}
\end{itemize}
i.e., the measurement matrix $\mathbf{A}_t$ behaves approximately as a tight frame with high Frobenius norm.

Given a measurement matrix $\mathbf{A} \in \mathbb{R}^{m \times n}$ that represents a sensor network of $m$ elements, our goal is to choose a subset of measurements $\mathbf{A}_t \in \mathbb{R}^{k_t \times n}$ from $\mathbf{A}$ such that we optimize the MSE$(\mathbf{A}_t)$, WCE$(\mathbf{A}_t)$ or the VCE$(\mathbf{A}_t)$ over all the time instances $t = 1,\dots,T,$ while we also balance the energy consumption of the network. 

\section{The sensor management problem}
\label{sec:background}

We defined now the sensor management problem for a single time instance, i.e., $T=1$. Given a network of $m$ sensors where each is capable of a linear measurement the sensor management problems asks which (and how many) sensors need to be activated in order to guarantee a fixed, given, performance measure (for example, the estimation accuracy in terms of MSE). An equivalent formulation can be made for example by fixing the number of sensors $k$ to activate while we minimize any of the performance measures \eqref{eq:mse}, \eqref{eq:wce} or \eqref{eq:vce}.

We consider that the full network of $m$ sensors is represented by $\mathbf{A} \in \mathbb{R}^{m \times n}$, i.e., each linear measurement is represented by a row $\mathbf{a}_i^T,\ i =1,\dots,m$. The selected sensors are denoted in the measurement matrix
\begin{equation}
	\mathbf{A}_1 = \begin{bmatrix} \mathbf{a}_{i_1} & \mathbf{a}_{i_2} & \dots & \mathbf{a}_{i_k} \end{bmatrix}^T \in \mathbb{R}^{k \times n},
	\label{eq:theB}
\end{equation}
a subset of rows of $\mathbf{A}$ indexed in the set $\mathcal{I} = \{i_1, \dots, i_k\}$ of size $k \geq n$, such its performance in terms of $\text{MSE}(\mathbf{A}_1)$ or $\text{WCE}(\mathbf{A}_1)$ is below a given threshold $\gamma$ or, alternatively, a maximum given number of sensors $k$ is activated. Notice that in order to optimize MSE$(\mathbf{A}_1)$, WCE$(\mathbf{A}_1)$ or VCE$(\mathbf{A}_1)$ we need to verify spectral properties of
\begin{equation}
	\mathbf{A}_1^T \mathbf{A}_1 = \sum_{i \in \mathcal{I}} \mathbf{a}_i \mathbf{a}_i^T = \mathbf{A}^T \text{diag}(\mathbf{z}) \mathbf{A} \in \mathbb{R}^{n \times n},
\end{equation}
where $\mathbf{z} \in \{0,1\}^n$ with $z_i = 1$ if $i \in \mathcal{I}$ and zero otherwise.

There are several approaches in the literature to deal with the sensor management problem. Although there are algorithms for to the sensor management problem that use search techniques \cite{Search1, Search2} or cross-entropy optimization \cite{Naeem2009}, we mainly distinguish two major approaches based on convex optimization and greedy methods and we discuss them separately in the next two subsections.

\subsection{Convex relaxation approach} In this formulation, the sensor selection problem is relaxed to a convex optimization \cite[Chapter~7.5]{CO}\cite{JoshiBoyd2009} program as
	\begin{equation}
		\begin{aligned}
			& \underset{\mathbf{z} \in [0,1]^m}{\text{maximize/minimize}} & & f(\mathbf{A}, \mathbf{z}) \\
			& \quad \quad \text{subject to} & & \mathbf{1}^T \mathbf{z} = k.
			\label{eq:solveconvex}
		\end{aligned}
	\end{equation}
	The goal is to construct a binary solution $\mathbf{z}$ that selects $k$ sensors such that $z_i$ indicates whether the $i^\text{th}$ sensor is selected or not. The objective function can be adapted to any of the performance measures in \eqref{eq:mse} \eqref{eq:wce} \eqref{eq:vce}. For example, when considering the VCE$(\mathbf{A}_1)$ we maximize $f(\mathbf{A}, \mathbf{z}) = \log\! \det (\mathbf{A}^T \text{diag}(\mathbf{z}) \mathbf{A})$ since the logarithm of the determinant is concave while for the MSE$(\mathbf{A}_1)$ we minimize $f(\mathbf{A}, \mathbf{z}) = \text{tr}((\mathbf{A}^T \text{diag}(\mathbf{z}) \mathbf{A})^{-1})$ since the trace of the inverse is convex. In order to reach a convex optimization problem the hard binary constraint $\mathbf{z} \in \{0,1\}^m$ is relaxed to $\mathbf{z} \in [0,1]^m$. Unfortunately, the problem in \eqref{eq:solveconvex} does not produce binary solutions $\mathbf{z}$ in general, just sub-unitary entries as the relaxation dictates. A rounding procedure, and usually also a local search, follow.
	
	A similar approach called SparSenSe is proposed in \cite{SparSenSe2014} where the same core optimization problem minimizes the MSE by selecting a few sensors given a maximum accepted level of MSE (not a fixed number of activated sensors). Again, a rounding procedure and potentially a local search can follow.

\subsection{Greedy methods approach}

The work of \cite{Vikalo2010} proposes to activate the sensors in the network according to a greedy procedure. For example, to minimize MSE in this fashion, given a measurement matrix $\mathbf{A}_1$ as in \eqref{eq:theB} a greedy scheme asks how to choose a new measurement $\mathbf{a}_j^T$ from the given full set $\mathbf{A} \in \mathbb{R}^{m \times n}$ such that 
\begin{equation}
	j = \underset{j \in \{1,\dots, m \} \setminus \mathcal{I}}{\arg \min} \text{ MSE}(\mathbf{\tilde{A}}_1),\text{ with } \mathbf{\tilde{A}}_1^T = \begin{bmatrix} \mathbf{A}_1^T & \mathbf{a}_j
	\end{bmatrix},
\end{equation}
and then adds the measurement to the active set
\begin{equation}
	\mathcal{I} \leftarrow \mathcal{I} \cup \{ j \}.
\end{equation}

Alternatively, FrameSense \cite{FrameSense2014} uses a greedy procedure to remove sensors from the network one-by-one such that the resulting remaining sensors have their frame potential maximized. We define the frame potential as $\text{FP}(\mathbf{A}) = \sum_{i=1}^m \sum_{j=1}^m (\mathbf{a}_i^T \mathbf{a}_j)^2 =  \| \mathbf{A}^T \mathbf{A} \|_F^2$ (which has been shown to achieve its minimum value exactly for $\alpha-$tight frames \cite{FP2003}, in this case its minimum value is $n \alpha^2$). Using the FP, the authors of \cite{FrameSense2014} are able to define a submodular objective function and therefore the greedy approach they deploy is well suited \cite{SubmodularBounds1978} for the optimization problem they propose. This approach is interesting because in general the performance indicators \eqref{eq:mse} and \eqref{eq:wce} are not submodular functions under the activation of new sensors and therefore greedy methods do not seem particularly well suited for the sensor management problem. The submodularity of \eqref{eq:vce} is exploited in \cite{Vikalo2010} where the set of active sensors is built up at each step. Building up the set of active sensors is also attractive from a computational perspective since in general we choose $k \ll m$ and therefore \cite{Vikalo2010} will perform less iterations than FrameSense \cite{FrameSense2014}.

Since monitoring spatial phenomena can be modeled in the context of Gaussian processes, a greedy method with submodularity properties was proposed in \cite{GPPlacement2008} to solve the sensor placement problem with near-optimal results. Using the same information theoretic approach the authors propose a branch-and-bound method that guarantees the construction of the optimal solution.
	
Another proposed greedy strategy is to add each sensor one by one according to the maximal projection onto the minimum eigenspace of a defined dual observation matrix \cite{MPME2016}. The method is computationally efficient and very successful in selecting the most informative sensors because it takes into account all eigenvalues of the observation matrix to encourage the two desirable properties \eqref{eq:invertible} and \eqref{eq:almosttight}. Greedy approaches to the sensor selection problem face some difficulties. For example, as pointed in \cite{MPME2016}, we have hard limitations for the eigenvalues of $\mathbf{A}^T\mathbf{A}$ when adding measurements one at a time.

\noindent \textbf{Result 1.} Given a positive semidefinite matrix $\mathbf{A}^T \mathbf{A}$ with eigenvalues $\lambda_1 \geq \dots \geq \lambda_n$ then $\mathbf{A}^T\mathbf{A} + \mathbf{aa}^T$ has eigenvalues $\mu_1 \geq \dots \geq \mu_n$ that have the interlacing property
\begin{equation}
\mu_1 \geq \lambda_1 \geq \mu_2 \geq \lambda_2 \geq \dots \geq \mu_n \geq \lambda_n.
\label{eq:interlacing}
\end{equation}
\noindent \textit{Proof.}  A proof is given in \cite[Chapter~4]{Horn}.$\hfill \blacksquare$

This is one of the reasons given in \cite{MPME2016} that the goal to increase all the eigenvalues $\lambda_i$ with each new measurement. Consider also the following two results.

\noindent \textbf{Result 2.} Assume $\mathbf{A}^T\mathbf{A}$ has $n$ eigenvalues $\lambda_1 \geq \dots \geq \lambda_n \geq 0$ in arithmetic progression, i.e, $\lambda_i = \lambda_1 + (i-1)r,\ i = 2,\dots,n$ with $r<0$, then the largest eigenvalue $\mu_1$ of $\mathbf{A}^T\mathbf{A} + \mathbf{aa}^T$, where $\mathbf{a} \in \mathbb{R}^n$ is a new measurement, obeys
\begin{equation}
\mu_1 \geq (\lambda_1 +(n-1)r) (1+\mathbf{a}^T (\mathbf{A}^T \mathbf{A})^{-1}\mathbf{a}).
\end{equation}

\noindent \textit{Proof.} See Appendix A. $\hfill \blacksquare$

\noindent \textbf{Result 3.} Given the measurements $\mathbf{A} \in \mathbb{R}^{m \times n}$ and any new measurement $\mathbf{a} \in \mathbb{R}^n$ then the new measurement matrix $\mathbf{\tilde{A}}_1^T = \begin{bmatrix} \mathbf{A}_1^T & \mathbf{a} \end{bmatrix} \in \mathbb{R}^{n \times (m+1)}$, improves all performance measures, i.e., $\text{MSE}(\mathbf{\tilde{A}}_1) \leq \text{MSE}(\mathbf{A}_1)$, $\text{WCE}(\mathbf{\tilde{A}}_1) \leq \text{WCE}(\mathbf{A}_1)$, $\text{VCE}(\mathbf{\tilde{A}}_1) \geq \text{VCE}(\mathbf{A}_1)$ and equality holds only when $\mathbf{a} = \mathbf{0}_{n \times 1}$. Equivalently, the performance measures are monotonically decreasing functions with the number of measurements.

\noindent \textit{Proof.} This is an immediate consequence of Result 1. See Appendix B for a quantitative analysis of the result.$\hfill \blacksquare$

Result 2 shows that as the eigenvalues become more concentrated (a highly desired property by \eqref{eq:almosttight}) they might exhibit a repelling property with regard to the largest eigenvalue when a new measurement is added. This is turn means that though $\mathbf{A}^T\mathbf{A}$ behaves as a tight frame we have that $\mathbf{A}^T \mathbf{A} + \mathbf{aa}^T$ may no longer behave as such.

Consider the following example. After selecting $k$ sensors assume that we have reached a best case scenario where all the eigenvalues of $\mathbf{A}^T \mathbf{A}$ are the same, i.e., $\lambda_i(\mathbf{A}^T \mathbf{A}) = \alpha, \forall i = 1,\dots,n$, but by \eqref{eq:tightperformance} the resulting mean squared error $\text{MSE}(\mathbf{A}) = n \alpha^{-1}$ is still above a threshold we, the user, prefer. This means that extra sensors need to be selected to further reduce the MSE (by Result 3). When the first extra sensor is activated, by Result 2, we have that $\lambda_i(\mathbf{A}^T \mathbf{A}) = \lambda_i(\mathbf{A}^T \mathbf{A} + \mathbf{aa}^T)$ for $i=2,\dots,n-1$, i.e., only the largest eigenvalue has increased. For the same reason, Result 2, this behavior generalizes to the case when more extra sensors are added: from the group of eigenvalues that are equal, only the largest one increases its value. This means that starting from the measurements $\mathbf{A}$ such that $\mathbf{A}^T \mathbf{A}$ is a tight frame, we would need to activate in the best case scenario at least $n-1$ extra sensors in order to guarantee that we construct new measurements $\mathbf{\tilde{A}}$ such that $\mathbf{\tilde{A}}^T \mathbf{\tilde{A}}$ is again a tight frame, with better MSE.

\section{The proposed optimization techniques for sensor selection}

Given a sensor network, we expect the best accuracy to be achieved if all its sensors are activated. Therefore, we will express the MSE performance of $\mathbf{A}_1$ relative to the overall performance of the full network $\mathbf{A}$.

Therefore, we introduce the reference (lowest MSE) performance of the full network as
\begin{equation}
	\gamma_0 = \text{tr}((\mathbf{A}^T \mathbf{A})^{-1}).
	\label{eq:gamma0}
\end{equation}
We will impose estimation accuracy levels $\gamma = \rho \gamma_0$ where $\rho \geq 1$, ensuring that we are solving optimization problem which are feasible. Naturally, with larger $\rho$ we will select fewer sensors from the network (allowing larger mean squared error) and vice-versa. When $\rho=1$ the only feasible solution is to select the full sensor network.% We next describe the proposed optimization methods for the minimization of the MSE.% (extensions to the WCE and the VCE follow immediately).
\begin{algorithm}[t]
	\caption{ \textbf{-- Sensor scheduling by $\ell_1$ minimization. } \newline \textbf{Input: }The sensing matrix of the network with $m$ sensors $\mathbf{A} \in \mathbb{R}^{m \times n}$, the total number of time instances $T$, the maximum allowed error $\rho > 1$, the regularization parameter $\lambda > 0$, the vector of sensing costs $\mathbf{s} \in \mathbb{R}^{m}_+$ and the matrix of communication costs $\mathbf{C} \in \mathbb{R}^{m \times m}_+$. \newline \textbf{Output: } The scheduling table $\mathbf{Z} \in \{ 0,1 \}^{m \times T}$ for the sensor activations at each time step.}
	\begin{algorithmic}
		\State \textbf{Initialization:}
		
		\State \quad \textbf{1. } Set initial weights $\mathbf{w}_t = \mathbf{1}_{m \times 1}$ and initial all-zero solution $\mathbf{z}_t = \mathbf{0}_{m \times 1}$ for $t = 1,\dots,T$, i.e., $\mathbf{Z} = \mathbf{0}_{m \times T}$.
		
		\State \quad \textbf{2. } Initialize sets $\mathcal{N} = \emptyset$ indexing sensors that are not selected and $\mathcal{K} = \emptyset$ indexing sensors that are selected.
		
		\State \quad \textbf{3. } Establish the best MSE performance $\gamma_0$ by \eqref{eq:gamma0}.
		
		\State \textbf{Iterations:}
		
		\State \quad \textbf{1. } Set $\mathbf{Z}^\text{(prev)} \leftarrow \mathbf{Z}$.
		
		\State \quad \textbf{2. } Update weights according to $w_{ij} = (z_{ij}^{\text{(prev)}} + \epsilon)^{-1}$.
		
		\State \quad \textbf{3. } Solve \eqref{eq:myoptproblemconvexovertimeimplicit} or \eqref{eq:myoptproblemconvexovertimeexplicit} with the $\rho$ provided and the additional linear equality constraints $z_{ij} = 1, \ \forall \ (i,j) \in \mathcal{K},$ and $z_{ij} = 0, \ \forall \ (i,j) \in \mathcal{N},$ to obtain the current estimate $\mathbf{Z}$ via a standard convex optimization solver \cite{CVX}.
		
		\State \quad \textbf{4. } Update the sets $\mathcal{N} = \{(i,j)\ |\ z_{ij} \leq \epsilon\}$ and $\mathcal{K} = \{(i,j)\ |\ z_{ij} \geq 1-\epsilon\}$.
		
		\State \quad \textbf{5. } If the IRL1 iterative process of \eqref{eq:myoptproblemconvexovertimeimplicit} or \eqref{eq:myoptproblemconvexovertimeexplicit} has converged (or 20 iterations have been completed since the last convergence) and the solution is not binary, i.e., $\| \mathbf{Z} - \mathbf{Z}^\text{(prev)} \|_F^2 \leq \epsilon$ and $|\mathcal{N}| + |\mathcal{K}| \neq mT$, then set $\mathcal{K} \leftarrow \mathcal{K} \cup \{ \underset{(i,j)}{\arg \max}
				\ z_{ij},\ (i,j) \notin \mathcal{K} \}$ and update $\mathbf{Z}$ such that $z_{ij} = 1,\ \forall \ (i,j) \in \mathcal{K}$.
		
		\State \quad \textbf{6. } If solution is binary, i.e., $|\mathcal{N}| + |\mathcal{K}| = mT$, then stop otherwise go to step 1 of the iterative process.
	
	\end{algorithmic}
\end{algorithm}

\subsection{Energy constraints over multiple time instances}

We assume we are given a network of sensors and the goal is to select the most informative subset of sensors from the network (i.e., the subset of sensors that achieves some level of accuracy or mean squared error). This formulation is equivalent to asking where sensors need to be placed (from a fixed set of possible locations) such that the resulting network achieves a minimum level of prescribed accuracy. 

In the previous formulation, a particular sensor in the network is either selected or not (or equivalently, we place a sensor in a particular place or not) for the whole lifetime of the network. In some situations this scenario is realistic while in others it may not be a proper approach. Consider for example a scenario where sensors are placed in an observation field where each sensor has its own energy supply and communications capabilities. If we choose the most informative sensors and disregard their other constraints we end up with a solution that will never activate certain sensors, which are wasted in the network.

In this section we also deal with a scenario where our goal is to schedule how sensor from a network are selected over multiple time instances such that at each time instant we guarantee a certain level of accuracy (e.g., the MSE is below a threshold) while we also balance the energy and communications constraints of the sensors.

We deal with $m$ sensors to be scheduled over $T$ time instances and therefore we define the binary scheduling table
\begin{equation}
	\mathbf{Z} = \begin{bmatrix}
		\mathbf{z}_1 & \mathbf{z}_2 & \dots & \mathbf{z}_T
	\end{bmatrix} \in \{0,1\}^{m \times T},
\end{equation}
and we denote the scheduler at time $t$ by $\mathbf{z}_t \in \{0,1\}^m$, i.e., the columns of $\mathbf{Z}$, we denote $z_{ij}$ the $(i,j)^\text{th}$ entry of $\mathbf{Z}$ and we denote $z_i$ the entries of $\mathbf{z}_t$. We next propose two ways to construct this scheduling table.

Implicit energy constraints can be used to ensure that over the $T$ time instances we do not selected the same sensors each time. Therefore, we propose the following regularized convex optimization problem
\begin{equation}
\begin{aligned}
&\underset{\mathbf{Z} \in [0,1]^{m \times T}}{\text{minimize}} & & \sum_{t = 1}^T \mathbf{w}_t^T \mathbf{z}_t + \lambda \max \left( \mathbf{W} \sum_{t=1}^T \mathbf{z}_t \right) \\
&\text{subject to} & & \text{tr}( (\mathbf{A}^T \text{diag}(\mathbf{z}_t) \mathbf{A})^{-1} ) \leq \rho \gamma_0 \\
& & & \sum_{t=1}^T \mathbf{z}_t \geq \mathbf{1}_{m \times 1},
\end{aligned}
\label{eq:myoptproblemconvexovertimeimplicit}
\end{equation}
where the weight vectors $\{ \mathbf{w}_t \}_{t=1}^T \in \mathbb{R}^{m}$ are given as $w_{ij} = (z_{ij} + \epsilon)^{-1},\ i=1,\dots,m,\ j=1,\dots,T$, the performance level $\rho > 1$ is given and fixed, $\mathbf{W} \in \mathbb{R}^{m \times m}$ is a diagonal matrix whose entries are positive and describe the cost of using any one sensor relative to the others. We also add an explicit constraint to ensure that each sensor is selected at least once over all $T$ time instances (if the power capacity of some sensors is low then they are not added to the optimization problem). The inequality in the second constraint applies elementwise. The regularization then penalizes the repeated use of the same sensor. For example, if sensor $i$ has no energy constraints then we set $W_{ii} = 0$ while if $W_{ii} = 1$ and $W_{jj} = 2$ is interpreted as the fact that the $j^\text{th}$ sensor has half of the energy supply of the $i^\text{th}$ sensor.

An explicit energy constraint can be used to formulate the same problem when energy profiles of the sensors are available. Consider for example the following regularized convex optimization problem
\begin{equation}
\begin{aligned}
&\underset{\mathbf{e} \geq \mathbf{0},\ \mathbf{Z} \in [0,1]^{m \times T}}{\text{minimize}} & & \sum_{t = 1}^T \mathbf{w}_t^T \mathbf{z}_t + \lambda g(\mathbf{e}) \\ % \max(\mathbf{e}) \\
&\text{\ \ \ subject to} & & \text{tr}( (\mathbf{A}^T \text{diag}(\mathbf{z}_t) \mathbf{A})^{-1} ) \leq \rho \gamma_0 \\
& & & \text{diag}(\mathbf{s})\mathbf{Z1}_{T \times 1} \leq \mathbf{e}_0 + \mathbf{e} \\
& & & \sum_{t=1}^T \mathbf{z}_t \geq \mathbf{1}_{m \times 1},
\end{aligned}
\label{eq:myoptproblemconvexovertimeexplicit}
\end{equation}
where $\mathbf{e}_0 \in \mathbb{R}^m$ with $(\mathbf{e}_0)_i$ denoting the reference amount of energy available to the $i^\text{th}$ sensor and $\mathbf{s} \in \mathbb{R}_+^{m}$ denoting the cost of using the $i^\text{th}$ sensor once (the cost of sensing and processing). The idea of introducing $\mathbf{e}_0$ is to establish a threshold for the energy consumption of each sensor. Ideally, the goal is to construct $\mathbf{Z}$ such that $\mathbf{e} = \mathbf{0}_{m \times 1}$, i.e., the sensor network can operate with the prescribed accuracy and energy consumption constraints, but we do not know a priori if this is possible (meaning that the problem could be unfeasible). Therefore, by regularizing with $g(\mathbf{e})$ the objective function penalizes any excess of energy consumption above the threshold $\mathbf{e}_0$.

The sensing cost vector $\mathbf{s}$ and the power upper reference values $\mathbf{e}_0$ are supplied by the user. Deviations from $\mathbf{e}_0$ are penalized in the proposed optimization problem. The regularization function $g(\mathbf{e})$ can be chosen to be either an $\ell_2$ penalty $g(\mathbf{e}) = \| \mathbf{e} \|_2$ or an $\ell_\infty$ penalty $g(\mathbf{e}) = \| \mathbf{e} \|_\infty$.

\begin{figure}[t]
	\centering
	\includegraphics[trim = 278 311 228 125, clip, width=0.42\textwidth]{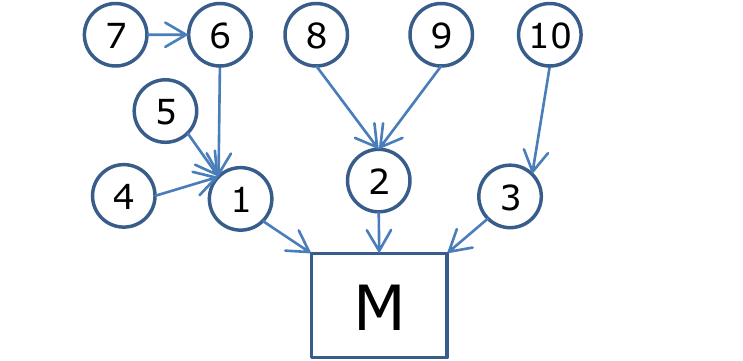}
	\caption{An example of network topology consisting of $m= 100$ sensors and a master node M. In the scenarios we consider each sensor has a dual role: it can perform a linear measurement and/or forward data from other sensors. To avoid routing issues, we consider a single (possible multi-hop) predefined path from each sensor to the master.}
	\label{fig:network}
\end{figure}

Finally, another energy constraint may deal with the cost of each sensor to transmit its measurement to a centralized node where the data from the full network is processed. Throughout this paper we have assumed this centralized model. In a simple scenario, each sensor is able to communicate directly to the central processing node. In this case, the cost of communication can be integrated into the cost of sensing denoted by $\mathbf{s}$ in \eqref{eq:myoptproblemconvexovertimeexplicit}. Otherwise, depending on the network topology, the energy constraint in \eqref{eq:myoptproblemconvexovertimeexplicit} can be modified to
\begin{equation}
	(\text{diag}(\mathbf{s}) + \mathbf{C}) \mathbf{Z1}_{T \times 1} \leq \mathbf{e}_0 + \mathbf{e}.
\end{equation}
This inequality applies elementwise. We have denoted by $\mathbf{C} \in \mathbb{R}_+^{m \times m}$ the communication cost of all the sensors. The entry $C_{ij} \geq 0$ expresses the cost incurred by the $i^\text{th}$ sensor in order to convey data from the $j^\text{th}$ sensor to the central node. When the $i^\text{th}$ sensor has a direct link to the central processing node then on the $i^\text{th}$ row of $\mathbf{C}$ has only one non-zero entry, namely $C_{ii}$. To clearly illustrate this constraint, an example network with nine sensors is given in Fig. \ref{fig:network} and its cost matrix is:
\begin{equation}
	\mathbf{C} = \begin{bmatrix}
					* & 0 & 0 & 0 & 0 & 0 & 0 & 0 & 0 & 0 \\
					0 & * & 0 & 0 & 0 & 0 & 0 & 0 & 0 & 0 \\
					0 & 0 & * & 0 & 0 & 0 & 0 & 0 & 0 & 0 \\
					* & 0 & 0 & * & 0 & 0 & 0 & 0 & 0 & 0 \\
					* & 0 & 0 & 0 & * & 0 & 0 & 0 & 0 & 0 \\
					* & 0 & 0 & 0 & 0 & * & 0 & 0 & 0 & 0 \\
					* & 0 & 0 & 0 & 0 & * & * & 0 & 0 & 0 \\
					0 & * & 0 & 0 & 0 & 0 & 0 & * & 0 & 0 \\
					0 & * & 0 & 0 & 0 & 0 & 0 & 0 & * & 0 \\
					0 & 0 & * & 0 & 0 & 0 & 0 & 0 & 0 & * \\
	\end{bmatrix} \in \mathbb{R}^{10 \times 10}.
\end{equation}
The non-zero values $C_{ij}$, shown here as ``*", are interpreted as the cost of transmitting data from the $i^\text{th}$ to the $j^\text{th}$ sensor. The $i^\text{th}$ line of $\mathbf{C}$ denotes the total cost of the sensor network to transmit data from the $i^\text{th}$ sensor to the master node. Most of the previous work (including the papers discussed in Section \ref{sec:background}) deals with a single time instance, i.e., $T=1$, in which case the proposed problem is equivalent to the reweighted $\ell_1$ approach used in \cite{ChepuriLeus2015} for non-linear measurement models.

Algorithm 1 can be directly extended to cover the WCE and VCE criteria. If we consider the WCE then the constraint becomes $\lambda_{\min} ( (\mathbf{A}^T \text{diag}(\mathbf{z}_t)\mathbf{A})^{-1} ) \leq \rho \gamma_0,\ \rho \geq 1$ where $\gamma_0 = \lambda_{\min}(\mathbf{A}^T \mathbf{A})$ and if we consider the VCE we have $\log \det (\mathbf{A}^T \text{diag}(\mathbf{z}_t) \mathbf{A} ) \geq \rho \gamma_0,\ 0 < \rho \leq 1$ where $\gamma_0 = \log \det(\mathbf{A}^T \mathbf{A})$.

\subsection{The proposed optimization procedure}

Based on the convex optimization problems described we now propose a method for the sensor selection problem. The full procedure is depicted in Algorithm 1.

The method keeps track of two sets $\mathcal{N}$ and $\mathcal{K}$ that contain indices of the solution $\mathbf{Z}$ that are set to zero and one, respectively. The procedure terminates only when each entry of $\mathbf{Z}$ has been allocated to either $\mathcal{N}$ or $\mathcal{K}$. Usually, convex optimization approaches to the sensor management problem relax the hard binary constraint and therefore the solutions are not binary in general. As such, a rounding procedure usually follows. In our case, the proposed method deals with the rounding problem internally in step 5 of Algorithm 1: if the algorithm has converged and the solution $\mathbf{Z}$ is not binary then take the largest entry from $\mathbf{Z}$ that is not in $\mathcal{K}$ and round it to one (and add it to $\mathcal{K}$). Because the IRL1 steps are not guaranteed to converge in general, we check if 20 iterations have passed since the last time the IRL1 optimization procedure converged to a non binary solution -- if this has happened then the rounding procedure of step 5 is applied (a new element is added to $\mathcal{K}$ and $\mathbf{Z}$ is updated) and the algorithm continues to step 6. The procedure then continues until the solution $\mathbf{Z}$ is binary. The strength of the method lies in the fact that the rounding procedure (allocating entries of $\mathbf{Z}$ to zero or one) is used only in two particular situations: the entries are very close to the extreme values or the optimization procedure converges before obtaining a binary solution, in which case we apply the rounding procedure to only one entry (the largest that is not already in $\mathcal{K}$). This means that almost all entries of $\mathbf{Z}$ are decided by the optimization procedure and not by the rounding (in stark contrast with the direct $\ell_1$ approach \cite{JoshiBoyd2009} for example that performs a single rounding step).

From a computational perspective, the proposed algorithm is numerically efficient, i.e., it is solved in polynomial time by off the shelf solvers \cite{CVX}. Moreover, as the sets $\mathcal{N}$ and $\mathcal{K}$ grow in size the optimization problems solved in step 3 of Algorithm 1 have fewer and fewer variables (from the initial $mT$) -- once an entry $(i,j)$ has been allocated to $\mathcal{K}$ or $\mathcal{N}$ it will remain there for the remainder of the iterative steps. The only parameter of Algorithm 1 is set to $\epsilon = 10^{-3}$.

\section{Analysis of sensor selection}

The proposed optimization problems are semidefinite programs (SDPs) with binary constraints used in an iterative fashion and therefore their analysis in terms of the optimality of the solution is difficult. In this section we explore several alternative ways to analyze (on average and worst case) the performance of sensor management solutions.

\subsection{Results for a tight sensor network}

For the purpose of understanding the behavior of the sensor selection problem, in this section we focus only on sensor networks that are characterized by tight measurement matrices $\mathbf{A}$. While it is true that in practice we do not generally deal with such frames, this choice is in fact the optimal, as discussed in Section II, and it appears naturally when considering sampling strategies of bandlimited graph signals \cite[Section III B]{EfficientSamplingSetSelection2016} when their defined shift operator is symmetric (for example, when we use the Laplacian).

\noindent \textbf{Result 4.} Assume we are given a sensor network represented by measurements in the $\alpha-$tight frame $\mathbf{A} \in \mathbb{R}^{m \times n}$.  Selecting a subset of $n \leq k \leq m$ sensor measurements from $\mathbf{A}$ which we denote $\mathbf{A}_1 = \{ \mathbf{a}_i^T \}_{i \in \mathcal{K}} \in \mathbb{R}^{k \times n}$ with $\mathbf{A}_1^T \mathbf{A}_1 = \sum_{i \in \mathcal{K}} \mathbf{a}_{i} \mathbf{a}_{i}^T$ where $\mathcal{K} = \{ i_1, \dots, i_k \}, |\mathcal{K}| = k$ then we have
\begin{equation}
	\begin{aligned}
&\mathbb{E}[\text{MSE}(\mathbf{A}_1)] \! \! = \! \! \frac{mn}{(k-n+1)\alpha}, \mathbb{E}[\text{WCE}(\mathbf{A}_1)] \! \! \geq \! \! \frac{m}{(k-n+1)\alpha}, \\ &\quad \quad \quad \quad \mathbb{E}[\text{VCE}(\mathbf{A}_1)] = \log \left( n! \left( \frac{\alpha}{m}\right)^n  {k \choose k-n} \right).
	\end{aligned}
	\label{eq:averageperformance}
\end{equation}
\noindent \textit{Proof.} See Appendix C.$\hfill \blacksquare$

\begin{figure}[t]
	\centering
	\includegraphics[trim = 12 7 30 20, clip, width=0.35\textwidth]{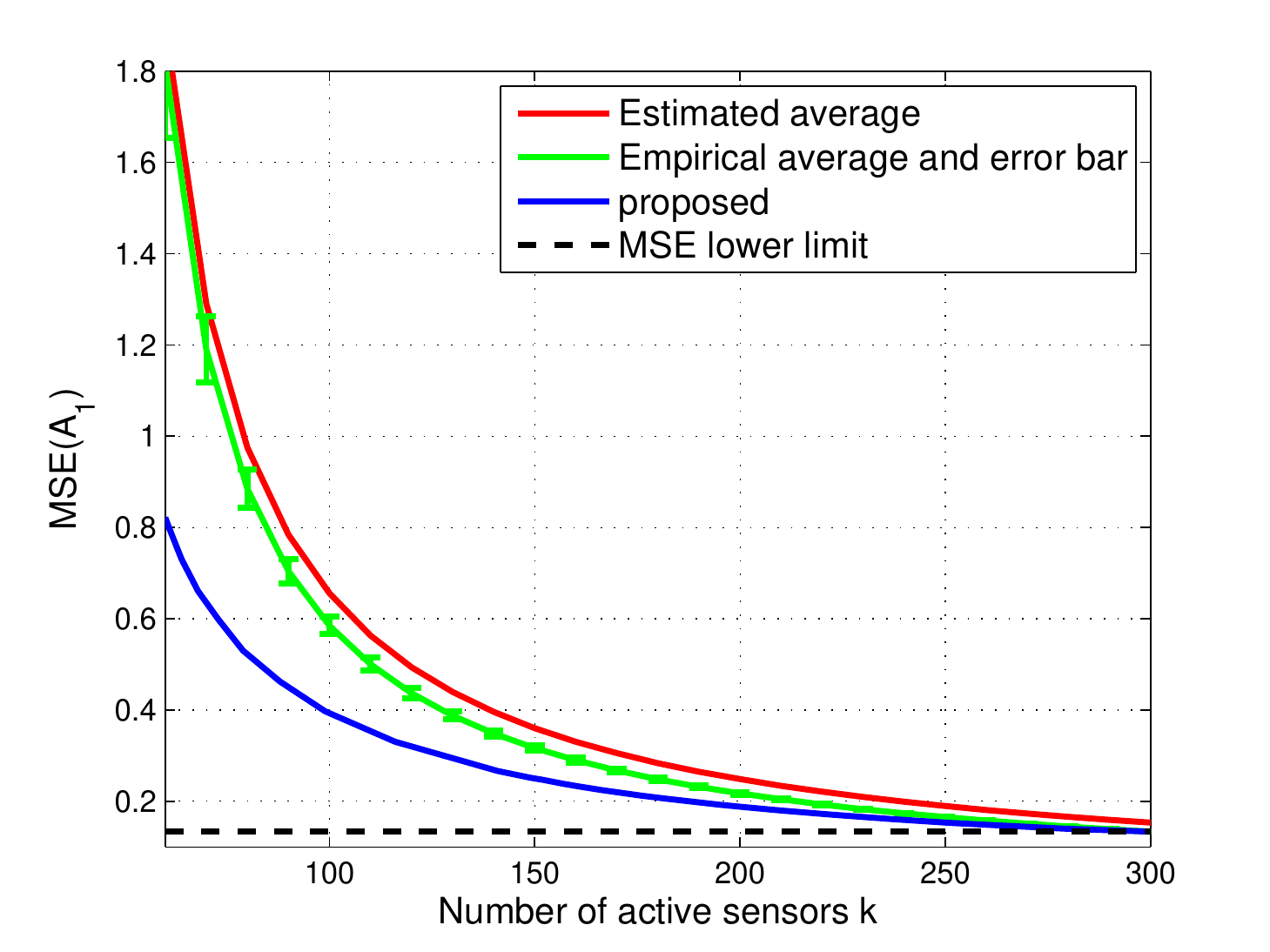}
	\caption{Expected versus empirical values of $\text{MSE}(\mathbf{A}_1)$ constructed by selecting $k$ sensors from a total of $m$ belonging to random tight frame $\mathbf{A} \in \mathbb{R}^{m \times n}$ with $\alpha = m = 300$ and $n = 40$. Empirical results are averaged over $10^5$ random sensor selections from $\mathbf{A}$. We show the lower limit of $\text{MSE}(\mathbf{A}_1)$ which is $\text{MSE}(\mathbf{A}) = n m^{-1}$ by \eqref{eq:tightperformance} and is achieved by $\mathbf{A}_1$ when $k = m$.}
	\label{fig:expectedMSE}
\end{figure}

This estimate is slightly biased. For example, when $k = m$ and $\alpha = m$ we know by \eqref{eq:tightperformance} that $\text{MSE}(\mathbf{A}_1) = \text{MSE}(\mathbf{A}) = nm^{-1}$ but we have the different estimate $\mathbb{E}[\text{MSE}(\mathbf{A}_1)] = n(m-n+1)^{-1} > \text{MSE}(\mathbf{A})$. Still, the gap between the two decreases as $m$ increases in a regime where $m \gg n$ -- in fact, the gap is approximately $(n/m)^{2}$. For some choice of $n,m$ and $\alpha$, this effect can be seen in Fig. \ref{fig:expectedMSE}: the estimated curve is above the one empirically observed. It is clear from the figure that the largest differences are for low $k$ (on the same order with $n$) while the gap closes for $k$ approaching $m$. As we will also see from the results section, the largest differences between the performance of the methods we analyze are for low values of $k$. Indeed, past research \cite{MPME2016} has shown by numerical experimentation that most of the sensor selection methods proposed in the literature perform similarly in the regime $k \gg n$.

Also, Result 4 shows that the MSE$(\mathbf{A}_1)$ and WCE$(\mathbf{A}_1)$ decrease on average linearly with the number of selected sensors. Dependencies with the other dimensions are also linear and intuitive: increasing the number of parameters to estimate ($n$) and the total number of available sensors ($m$) leads to worse performance; increasing the energy, essentially the signal to noise ratio, of the measurement matrix ($\alpha$) improves performance. Furthermore, the empirical standard deviation around the mean MSE decreases with $k$ showing that the largest potential gains in MSE can be achieved only in the regime where $k \approx n$.

\subsection{Relating sensor management to other problems}

In this subsection we connect the sensor selection problem to other fields of research. Subset selection problems have been seen in many areas of research. For example, the problem of selecting a subset of column from a matrix such that some spectral guarantees are obeyed is well studied. Topics such as column subset selection \cite{SubsetSelection2010} and the restricted invertibility problem \cite{RestrictedInvertability2012} deal with constructing a matrix by selecting a subset of columns from a given matrix such that the new construction has the lowest singular value well above zero (i.e., the new matrix is well conditioned).

We next detail some research topics closely related to the sensor selection problem and discuss how these results apply here and to previous empirical observations from the literature.

\noindent \textbf{Techniques that minimize the condition number of a matrix }\cite{MinCondNumber1} \cite{MinCondNumber2} can also be deployed for the sensor management problem. Given constants $s_0, t_0 \in \mathbb{R}_+$, consider the following semidefinite program:
\begin{equation}
\begin{aligned}
&\underset{t,s,\ \mathbf{z} \in [0,1]^m}{\text{minimize}} & & t-s + \lambda  \mathbf{w}^T \mathbf{z} \\
&\text{subject to} & & s\mathbf{I} \preceq \mathbf{A}^T \text{diag}(\mathbf{z}) \mathbf{A} \preceq t\mathbf{I} \\
& & &  s_0 \leq s \leq t \leq t_0.
\end{aligned}
\label{eq:myoptproblemconvexrcond}
\end{equation}
The inequality $s\mathbf{I} \preceq \mathbf{A}^T \text{diag}(\mathbf{z}) \mathbf{A}$ reads as $\mathbf{A}^T \text{diag}(\mathbf{z}) \mathbf{A} - s\mathbf{I}$ is positive semidefinite \cite[Chapter~2.2.5]{CO}. Sensor selection solutions provided by this optimization problem are well suited for our purposes since the constraints lead to the design of well invertible $ \mathbf{A}_1 = \mathbf{A}^T \text{diag}(\mathbf{z}) \mathbf{A}$ due to the threshold provided by the $s_0 \gg 0$ and a tight structure by the variables $s$ and $t$. Therefore, the two desired properties are enforced \eqref{eq:invertible} \eqref{eq:almosttight}. Solving the problem in \eqref{eq:myoptproblemconvexrcond} (assuming also some rounding procedure to construct a binary solution $\mathbf{z}$) gives explicit bounds on the eigenvalues of $\mathbf{A}_1$ and therefore by \eqref{eq:mse} and \eqref{eq:wce} we have the following error bounds
\begin{equation}
	\frac{n}{t_0} \leq \text{MSE}(\mathbf{A}_1)  \leq \frac{n}{s_0},\ \frac{1}{t_0} \leq \text{WCE}(\mathbf{A}_1) \leq \frac{1}{s_0}.
\end{equation}
The upper bounds hold even after a rounding procedure is applied on the solution $\mathbf{z}$ of \eqref{eq:myoptproblemconvexrcond}.

\noindent \textbf{Compressed sensing }\cite{CS} makes use of measurement matrices $\mathbf{A} \in \mathbb{R}^{m \times n}, m>n,$ that obey the restricted isometry property
\begin{equation}
	(1-\delta_k)\|\mathbf{x}\|_2^2 \leq \|\mathbf{A} \mathbf{x} \|_2^2  \leq (1+\delta_k)\| \mathbf{x} \|_2^2,
	\label{eq:rip}
\end{equation}
for a constant $\delta_k$ and any $k-$sparse vector $\mathbf{x} \in \mathbb{R}^n$. Let us denote by $\mathbf{A}_1 \in \mathbb{R}^{k \times n}$ any subset of $k \geq n$ rows from $\mathbf{A}$. % Then, equivalently to \eqref{eq:rip} we have
%\begin{equation}
%	1-\delta_k \leq \lambda_n(\mathbf{A}_1^T \mathbf{A}_1) \leq \lambda_1(\mathbf{A}_1^T \mathbf{A}_1) \leq 1+\delta_k.
%	\label{eq:ripeigenvaluebound}
%\end{equation}
In the case of random matrices obtained from the standard Gaussian distribution $\mathcal{N}(0,1)$ we have by Gordon's theorem for Gaussian matrices \cite[Chapter~5]{CSBook} that
\begin{equation}
	(\sqrt{k} - \sqrt{n})^2 \! \leq \! \lambda_n(\mathbf{A}_1^T \mathbf{A}_1) \! \leq \! \lambda_1(\mathbf{A}_1^T \mathbf{A}_1) \! \leq \! (\sqrt{k} + \sqrt{n})^2.
\end{equation}
%In the case of random matrices obtained from a Gaussian distribution $\mathcal{N}(0, 1/n)$ we have that $\delta_k \approx 2\sqrt{kn^{-1}} + kn^{-1}$. Unfortunately, these bounds \eqref{eq:ripeigenvaluebound} on the eigenvalues hold when $k/n \to 0$ while for the sensor management problem we are interested in the regime $k\geq n$. When $k \geq n$ we still have
%\begin{equation}
%	\lambda_1(\mathbf{A}_1 \mathbf{A}_1^T) \leq (1+\delta_k).
%\end{equation}
Having these bounds on the extreme eigenvalues of $\mathbf{A}_1^T \mathbf{A}_1$, it follows directly that
\begin{equation}
	\frac{n}{ (\sqrt{k} + \sqrt{n})^2} \leq \text{MSE}(\mathbf{A}_1) \leq \frac{n}{ (\sqrt{k} - \sqrt{n})^2}.
	\label{eq:randommse}
\end{equation}
This is similar to the simple bound in \eqref{eq:tightperformance} since $\mathbb{E}[\| \mathbf{A}_1 \|_F^2] = kn$ and therefore $\mathbb{E}[\text{MSE}(\mathbf{A}_1)] \geq n/k$. Notice that in \eqref{eq:randommse} when $k \gg n$ we have that $\text{MSE}(\mathbf{A}_1) \approx n/k$ and therefore, asymptotically as $k \to m$, $\mathbf{A}_1$ behaves as an $\alpha-$tight frame with $\alpha = 1$ (the tightness of random frames was established by \cite[Theorem~1]{goyal1998quantized}). Other results from non-asymptotic random matrix theory \cite[Chapter~5]{CSBook} can also be used here to understand behavior of the extremal eigenvalues and bound the MSE with high probability.

\noindent \textbf{The solution to the Kadison-Singer problem }\cite{KadisonSinger2015} shows that given a tight $\mathbf{A} \in \mathbb{R}^{m \times n}$, i.e., $\alpha = 1$, where $\delta = \underset{i = 1,\dots,m}{\max} \| \mathbf{a}_i^T \|_2^2$ is the maximum squared $\ell_2$ norm of the rows of $\mathbf{A}$ there exists a partition of the rows into $T$ sets $\{S_1, \dots, S_T\}$ such that
\begin{equation}
	\! \! \! \! \! \sigma_1(\mathbf{A}_{S_t}) \! \! = \! \! \sqrt{\lambda_1(\mathbf{A}_{S_t}^T \mathbf{A}_{S_t})} \! \leq \! \left( \frac{1}{\sqrt{T}} + \sqrt{\delta} \right)^2\! \! \!,t=1,\dots,T.
	\label{eq:sigma1}
\end{equation}
Therefore with $T = m/k$ for a tight $\mathbf{A}$ and denoting $\mathbf{A}_t = \mathbf{A}_{S_t} \in \mathbb{R}^{|S_t| \times n}$ the matrix composed of the rows from $\mathbf{A}$ indexed in the set $S_t$, with $|S_t| = k$, we have
\begin{equation}
	\lambda_1(\mathbf{A}_t^T \mathbf{A}_t ) \leq \left( \sqrt{\frac{k}{m}} + \sqrt{\delta} \right)^4,\ t=1,\dots,T.
	\label{eq:lambda1}
\end{equation}
The value $\delta$ for any measurement matrix $\mathbf{A}_t \in \mathbb{R}^{k \times n}$ can be estimated by using the Markov inequality and a union bound to show that
\begin{equation}
	\begin{aligned}
	\mathbb{P} \left( \underset{i=1,\dots,k}{\max} \right. &  \| \mathbf{a}_i^T \|_2^2 \geq c \mathbb{E}[\| \mathbf{a}_i^T \|_2^2] \bigg) \leq \\
	& \sum_{i=1}^k  \mathbb{P}( \| \mathbf{a}_i^T \|_2^2 \geq c \mathbb{E}[\| \mathbf{a}_i^T \|_2^2]) \leq \frac{k}{c},
	\end{aligned}
\end{equation}
where $\mathbb{E}[\| \mathbf{a}_i^T \|_2^2] = n m^{-1}$ and $c = c_1 k, \ c_1 \geq 1$. Therefore, with high probability $1-c_1^{-1}$ we have that $\delta < c_1 k n m^{-1}$. Still, empirically we observe that $\delta$ is within a constant factor of the expected value $nm^{-1}$, i.e., that $c = O(1)$, not $c = O(k)$.

Therefore, all measurement matrices $\mathbf{A}_t$ at each time instance $t = 1,\dots,T,$ obey
\begin{equation}
\begin{aligned}
\text{MSE}&(\mathbf{A}_t) =  \sum_{i=1}^n \frac{1}{\lambda_i(\mathbf{A}_t^T\mathbf{A}_t)}
\geq \frac{n}{\lambda_1(\mathbf{A}_t^T\mathbf{A}_t)} \\
\geq & \frac{n m^2}{k^2} \left( 1 + \sqrt{ \frac{\delta m}{k}} \right)^{-4} \\
= & \frac{n m^2}{(\sqrt{k} + \sqrt{c_1 n})^4} \approx
\begin{cases}
n \left( \frac{m}{k} \right)^2,      & \text{if } k \gg n \\
\frac{m}{(1+\sqrt{c_1})^4} \frac{m}{k},  & \text{if }k \approx n.\\
\end{cases}
\end{aligned}
	\label{eq:lowerbound}
\end{equation}
This result shows that potentially the MSE can exhibit a quadratic decrease with the number of selected sensors $k$. This is to be compared with \eqref{eq:averageperformance} that shows a linear decrease of the expected MSE with $k$. When $k = m$ the bound in \eqref{eq:lowerbound} matches the optimal value of MSE in \eqref{eq:tightperformance} for $\alpha = 1$. These bounds are also reflected in the results from Fig. \ref{fig:expectedMSE} where we can observe that for $k \approx n$ the decrease in MSE achieved by the proposed method, with the increased number of selected sensors $k$, is larger that of a random sensor selection algorithm when $k \gg n$. These insights confirm previous experimental results from the literature, like \cite{MPME2016}, where the methods proposed for sensors selection differ mostly when $k \approx n$ and are similar when $k \gg n$ (or $k \approx m$) where even random selections provide good estimation accuracy (low MSE and WCE).

It is important to mention that constructing a set $S_t$ such that \eqref{eq:sigma1} is always obeyed is still an open problem. Heuristics can be proposed similar to the approach presented in this paper for binary optimization (especially because the bound in \eqref{eq:lambda1} is a convex constraint).

%The solution to the Kadison-Singer problem is also useful in our discussion of operating a sensor network over multiple time instances.
Result \eqref{eq:sigma1} essentially states that a tight measurement matrix $\mathbf{A}$ can be partitioned such that the $T$ partitions $\mathbf{A}_t$ themselves are also approximately tight, i.e., all the $\mathbf{A}_t, \ t = 1,\dots,T,$ obey \eqref{eq:lowerbound}. For example, this links with our objective in \eqref{eq:myoptproblemconvexovertimeimplicit} of avoiding selecting the same sensors by partitioning the sensor set into disjoint subsets (a severe constraint) while ensuring that each subset still behaves well, i.e., similar estimation accuracy of the subsets according to MSE.

\begin{figure}[t]
	\centering
	\includegraphics[trim = 10 5 25 15, clip, width=0.35\textwidth]{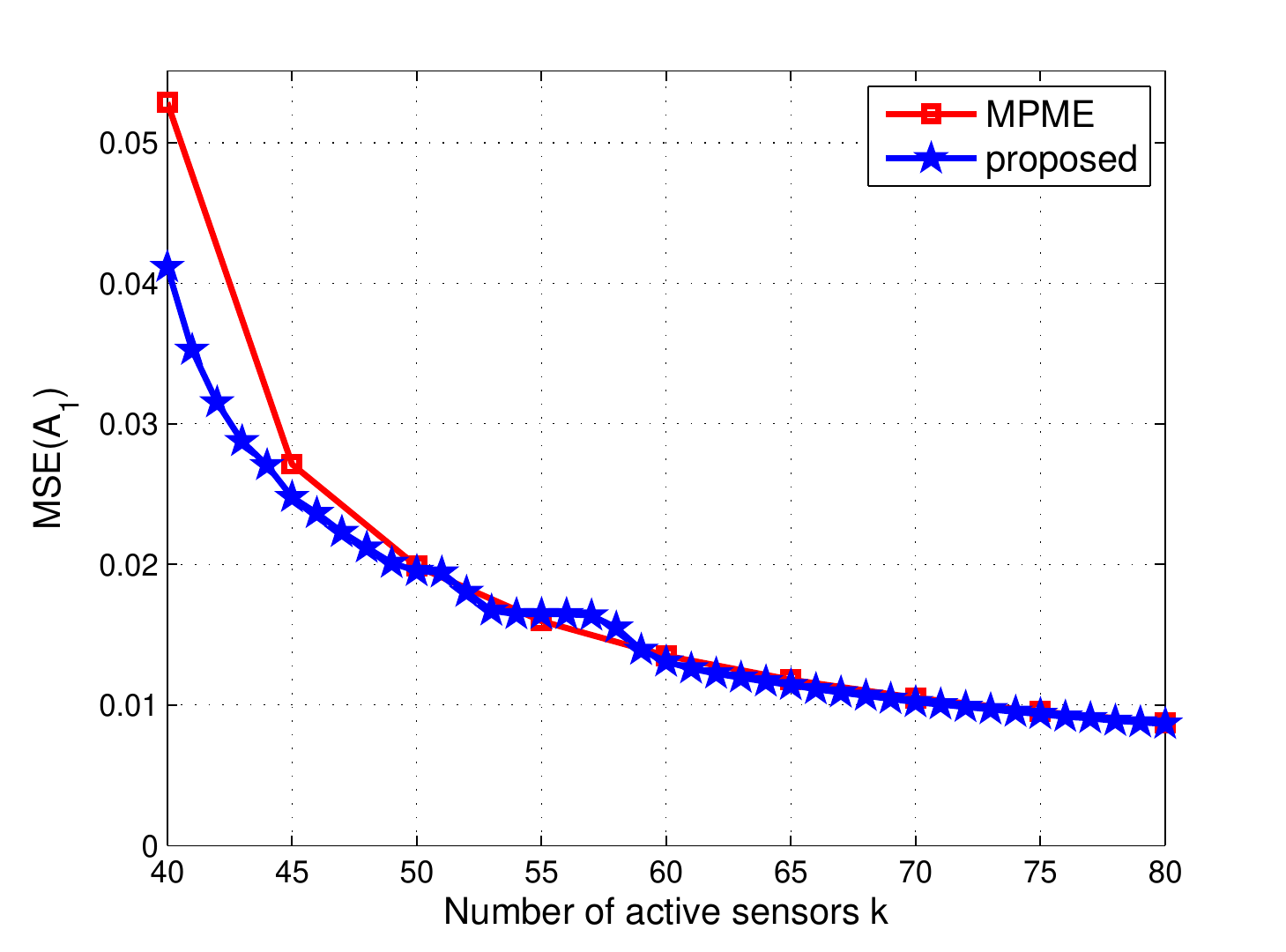}
	\caption{Average mean squared error reached by MPME \cite{MPME2016} and the proposed method for the estimation of a variable of size $n=40$ with a sensor network of maximum $m = 100$ elements. The measurement matrix is random Gaussian and the results shown are averaged over 100 realizations.}
	\label{fig:figure1}
\end{figure}
%\begin{figure}[t]
%	\centering
%	\includegraphics[trim = 3 5 25 15, clip, width=0.26\textwidth]{figs/figure2.eps}
%	\caption{Average worst case error reached by MPME \cite{MPME2016} and the proposed method for the same experimental setup as Fig. \ref{fig:figure1}.}
%	\label{fig:figure2}
%\end{figure}
\begin{figure}[t]
	\centering
	\includegraphics[trim = 10 5 25 15, clip, width=0.35\textwidth]{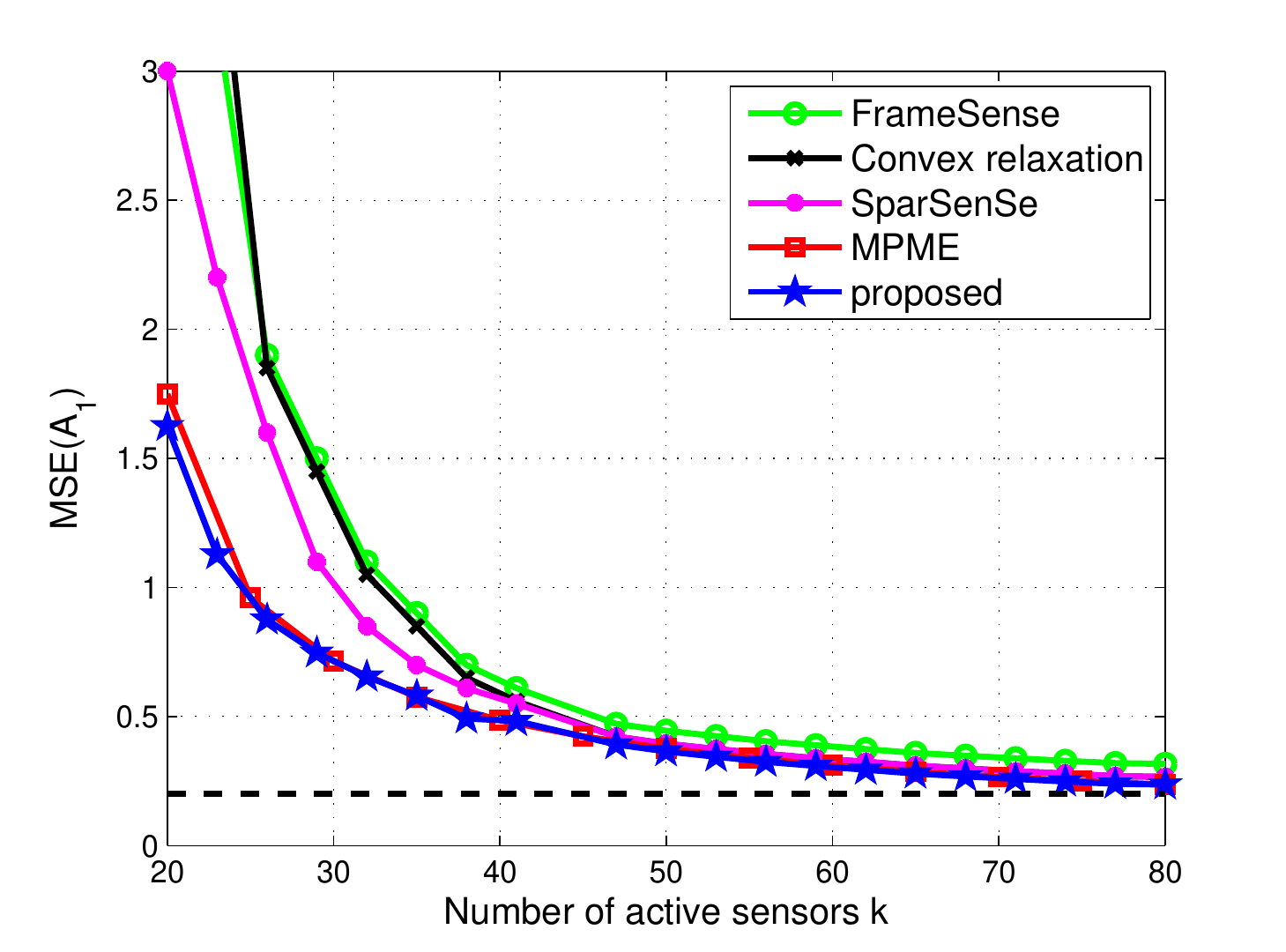}
	\caption{Comparison of average mean squared error for the estimation of a variable of size $n=20$ with a sensor network of $m = 100$ sensors. The measurement matrix is a random $\alpha-$tight with $\alpha = 100$ and the results are averaged over 100 realizations. We explicitly show the MSE lower bound value $n \alpha^{-1} = 0.2$.}
	\label{fig:figure3}
\end{figure}
%\begin{figure}[t]
%	\centering
%	\includegraphics[trim = 10 5 25 15, clip, width=0.26\textwidth]{figs/figure4.eps}
%	\caption{Comparison of average worst case error for the same experimental setup as Fig. \ref{fig:figure3}. We explicitly show the WCE lower bound value $ \alpha^{-1} = 0.01$.}
%	\label{fig:figure4}
%\end{figure}
\section{Results}

In this section we provide experimental numerical simulations to show the performance of the proposed methods and how they compare with state of the art approaches from the literature. We also present extensive numerical simulations to describe the performance of the proposed method to schedule a sensor network over time while also balancing power consumption.

\subsection{Choosing how many sensors to activate}

In the first experimental setting we provide numerical evidence on how the MSE evolves with the number of selected sensors given a fixed network. In this subsection we consider a sensor network whose measurements $\mathbf{A} \in \mathbb{R}^{m \times n}$ are described by a matrix with entries scaled i.i.d. random Gaussian from a zero mean distribution with variance one, i.e, $A_{ij} \sim \sqrt{m} \mathcal{N}(0, 1),\ i = 1,\dots,m,\ j = 1,\dots,n$.

Results for a network of $m = 100$ sensors tasked to estimate an unknown of size $n = 40$ are shown in Fig. \ref{fig:figure1}. We compare our proposed method with the state of the art approach MPME \cite{MPME2016}. As previously noted by empirical simulations \cite{MPME2016} the performance of the two methods in terms of MSE is similar when $k \gg n$ while there is a small gap in performance favoring the proposed method when $k \approx n$. These results are expected in light of the discussion in Section V, around Fig. \ref{fig:expectedMSE}, where we give theoretical arguments as to the difference between the $k \gg n$ and $k \approx n$ sensor selection regimes. To show the performance of the proposed method we evaluate it for $\rho \in (1, 10]$ on a fine grid. Fig. \ref{fig:figure1} provides an empirical practical way of choosing the number of sensor to activate while also balancing the MSE level. Up to $k \approx 60$ sensors, the MSE shows significant decrease while after this level each new sensor activation has important diminishing returns. Also, to approach the performance of the full network a large number (close to $m$) of sensors need to be activated.

\subsection{Comparisons with previous sensor selection algorithms}

Following the experimental setup from \cite{MPME2016}, in this section we compare the proposed method with previously proposed methods from the literature. We choose to simulate a sensor network with $m = 100$ elements tasked to recover an unknown of size $n=20$. Fig. \ref{fig:figure3} shows the simulation results where we compare with FrameSense \cite{FrameSense2014}, convex relaxation ($\ell_1$ followed by rounding, using the log determinant approach to minimize VCE) \cite{JoshiBoyd2009}, SparSenSe \cite{SparSenSe2014} and MPME \cite{MPME2016}. All measurement matrices used here are $\alpha-$tight with $\alpha = 100$. They were constructed after projecting random Gaussian matrices on the set of tight matrices (numerically this is done by taking the polar factor via a singular value decomposition: $\mathbf{A} \leftarrow \sqrt{\alpha} \mathbf{UV}^T$ from the random matrix $\mathbf{A} = \mathbf{U\Sigma V}^T$).

MPME \cite{MPME2016} and the proposed method perform best, with similar results. Since MPME \cite{MPME2016} is reported to outperform a direct, one shot, $\ell_1$ approach followed by rounding, these results show the benefit of using the iterative reweighted approach in Algorithm 1. Just as in the previous section, the proposed method seems to outperform MPME slightly when the number of selected sensors $k$ is close to $n$. The other methods perform significantly worse in this regime while the performance gaps mostly vanish when $k \gg n$. An exception to this observation is FrameSense \cite{FrameSense2014}, which exhibits higher MSE even when increasing $k$.

\begin{figure}[t]
	\centering
	\includegraphics[width=0.45\textwidth]{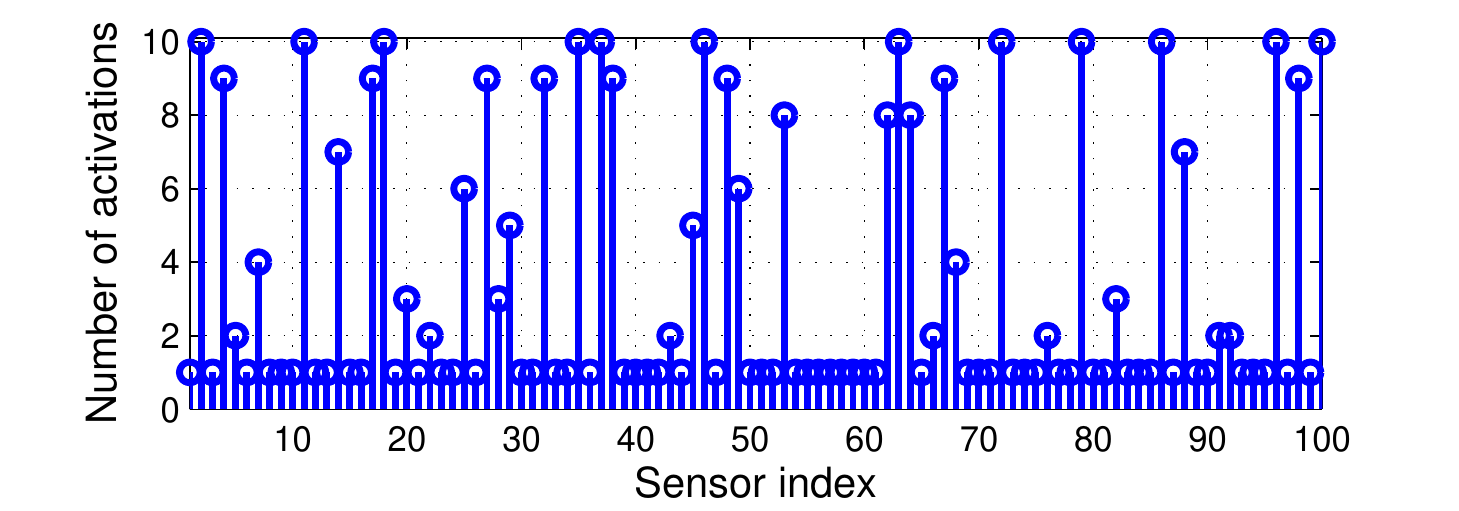}
	\caption{Total number of activation of each sensor from the $m = 100$ elements of a sensing network described by a $\alpha-$tight measurement matrix of size $100 \times 20$ with $\alpha = 100$. The implicit energy constraint \eqref{eq:myoptproblemconvexovertimeimplicit} runs with the regularization parameter $\lambda = 1$ and the optimization takes place of $T = 10$ time instances and the estimation accuracy is fixed to $\rho = 3$. Overall there and 342 sensors activations in the network.}
	\label{fig:figure5}
\end{figure}
\begin{figure}[t]
	\centering
	\includegraphics[width=0.45\textwidth]{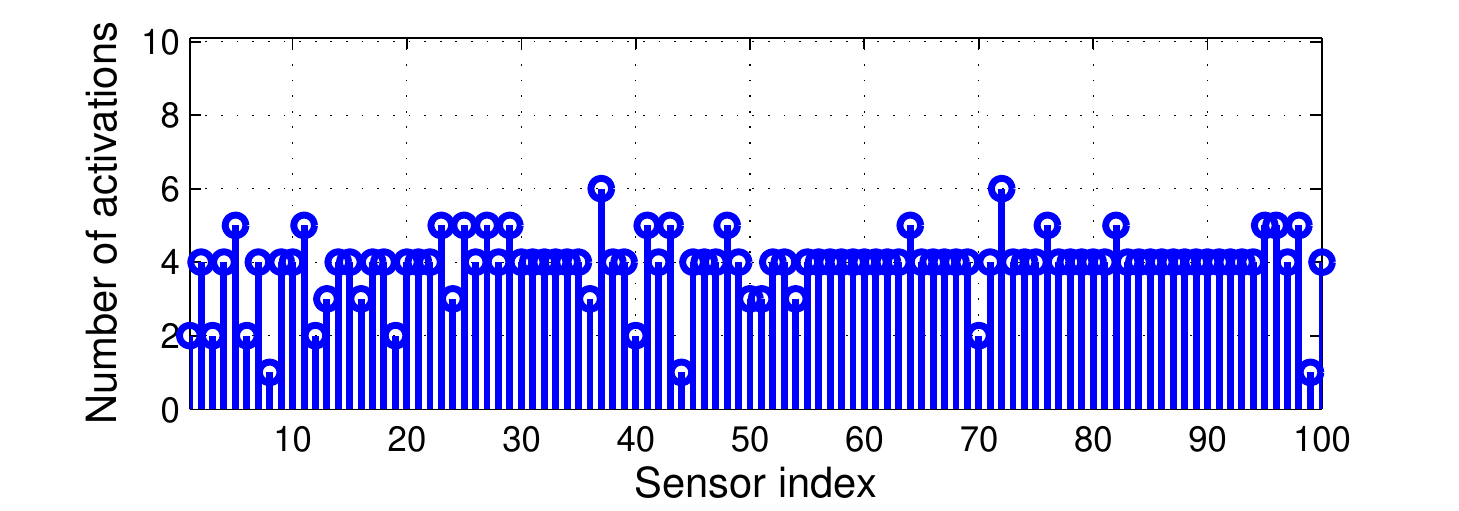}
	\caption{Total number of activation of each sensor for the same experimental setup as in Fig. \ref{fig:figure5} with the regularization parameter $\lambda = 100$. Overall there and 389 sensors activations in the network.}
	\label{fig:figure6}
\end{figure}
For similar MSE the computational complexities of the methods play an important role. Although the propose method runs in polynomial time (due to interior point solver like \cite{CVX}) greedy methods are in general preferable in terms of computational complexity. For the simulations we consider in this section, Algorithm 1 ran on average for 25 iterations. As shown in \cite{MPME2016}, the MPME method is best in terms of computational complexity. Still, convex optimization approaches have an edge when some extra constraints are added to the sensor selection problem (like, as we will see in the following section, energy and communications constraints). Also, all methods in this paper run without the local search mechanisms often deployed, like in \cite{MPME2016} or \cite{JoshiBoyd2009}, since these do not supply in general any great improvement.

\begin{figure}[t]
	\centering
	\includegraphics[width=0.45\textwidth]{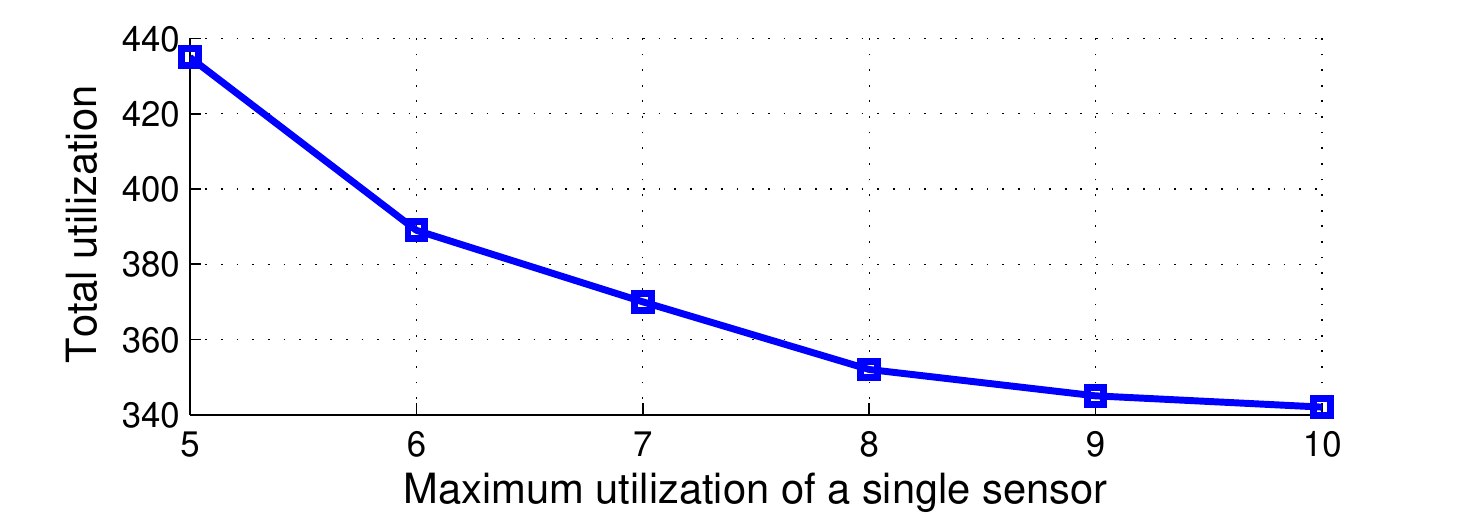}
	\caption{Maximum sensor utilization $\max \left( \sum_{t=1}^T \mathbf{z}_t \right)$ versus total sensor utilization $\sum_{i=1}^m \sum_{j=1}^T z_{ij}$ for a sensor network of $m = 100$ elements over $T = 10$ time instances. The experimental setup is such that in all six points of the plot we have the same minimum accuracy (MSE) in an estimation task performed by the network. The point $(10, 342)$ corresponds to Fig. \ref{fig:figure5} and the point $(6, 389)$ to Fig. \ref{fig:figure6}. The experimental setup is the same as in \ref{fig:figure5} but with various values for the regularization parameter $\lambda$.}
	\label{fig:figure8}
\end{figure}
\subsection{Sensor selection with energy and communication constraints}

To show the versatility of the convex optimization approach to the sensor selection problem we show how to deal with energy and communication constraints when scheduling the usage of a sensor network over multiple time instances.

First, we show the implicit energy constraint approach, i.e., without any explicit information about the energy profiles of the sensors our goal is to operate the sensor network over $T$ time instances such that we do not activate the same sensors at each time. The $\ell_1/\ell_\infty$ style optimization problem \eqref{eq:myoptproblemconvexovertimeimplicit} balances between the estimation accuracy of the sensor network and making sure that the sensing is distributed more evenly between the network's sensors. Results are shown in Fig. \ref{fig:figure5} and \ref{fig:figure6}. With the larger regularization parameter $\lambda = 100$ the results in Fig. \ref{fig:figure6} show a more balanced activation of the sensors, as opposed to the results in Fig. \ref{fig:figure5} that are obtained for a smaller $\lambda = 1$. With higher parameters $\lambda$ the sensor scheduling problem is regularized to select less often the same sensors (for example in Fig. \ref{fig:figure6} most sensors are selected four, five or maximum six times as compared to Fig. \ref{fig:figure5} where several sensors are selected in all ten time instances), but at the cost of activating, overall, a larger number of sensors over the ten time instances.

\begin{figure}[t]
	\centering
	\includegraphics[width=0.43\textwidth]{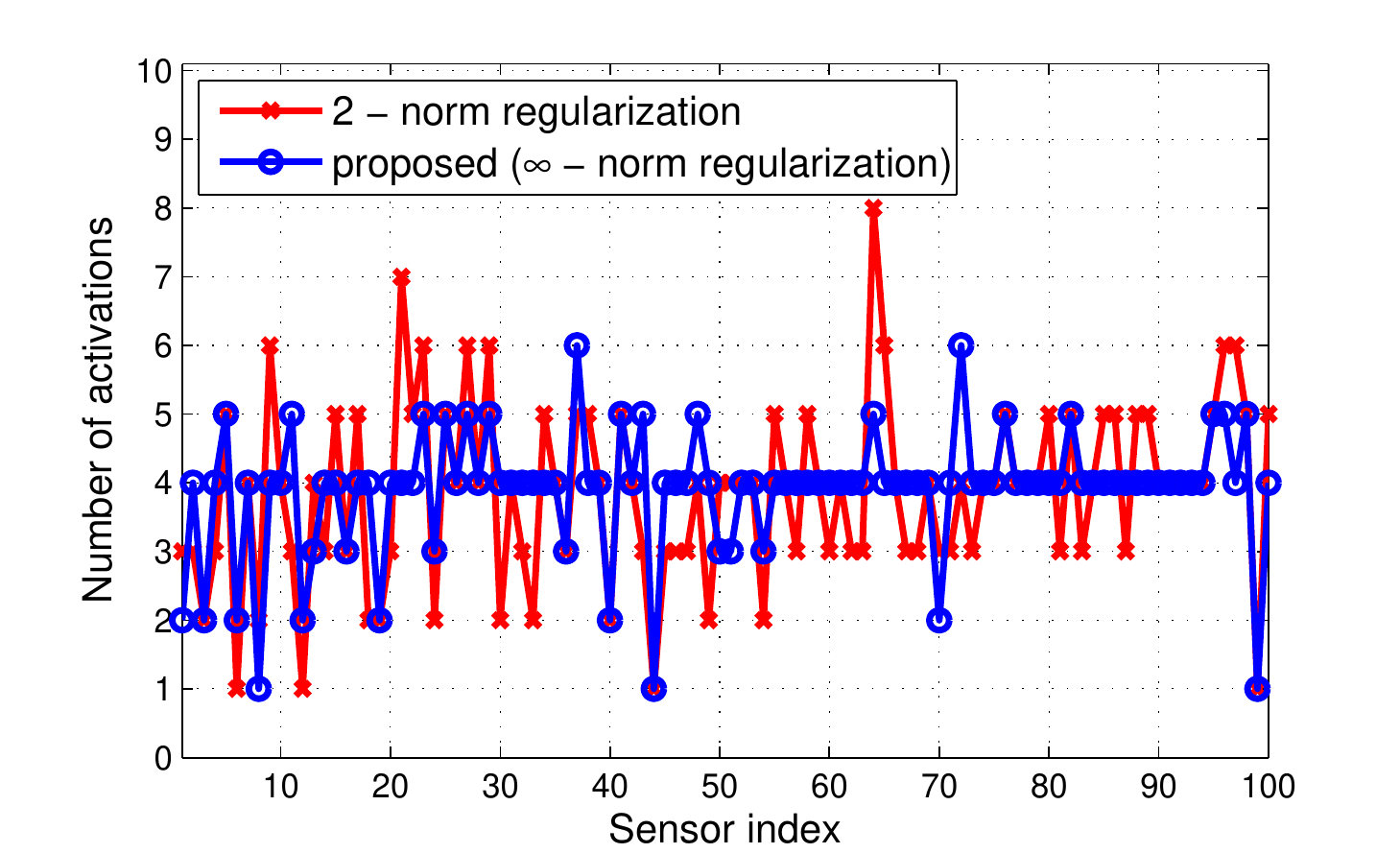}
	\caption{Comparison of the total number of activation of each sensor between the proposed approach and a previously introduced $\ell_2$ regularization approach \cite{LiuChenVempatyFardadShenVarshney2015} for selection balancing. The effect of the proposed $\ell_\infty$ regularization can be seen as most sensors are selected four times and the most used sensors peak at six. We have the same experimental setup as in Fig. \ref{fig:figure5}.}
	\label{fig:figure9}
\end{figure}

The almost flat envelope of Fig. \ref{fig:figure6} is typical of solutions to convex optimization problems that involve $\ell_\infty$ regularized objective functions (for details see \cite[Chapter~6]{CO}). The almost uniform activation of the sensors over time distributes the sensing workload of the network ensuring balanced power consumption together with increased robustness and fault tolerance in case of any particular sensor failure. Fig. \ref{fig:figure8} shows one of the side effects of the proposed optimization procedure: we can reduce the frequency with which one particular sensor is activated but at the cost of activating other (possibly many more) sensors from the network such that it operates with the same estimation accuracy. Fig. \ref{fig:figure9} compares the proposed algorithm with a previously introduced method that uses a $\ell_2$ regularized penalty \cite{LiuChenVempatyFardadShenVarshney2015}, instead of the $\ell_\infty$ used in this paper, to discourage the selection of the same sensors over time. Our method shows a lower variance and a lower maximum for the total number of activations of each separate sensor.

\begin{figure}[t]
	\centering
	\includegraphics[trim = 10 5 25 5, clip, width=0.35\textwidth]{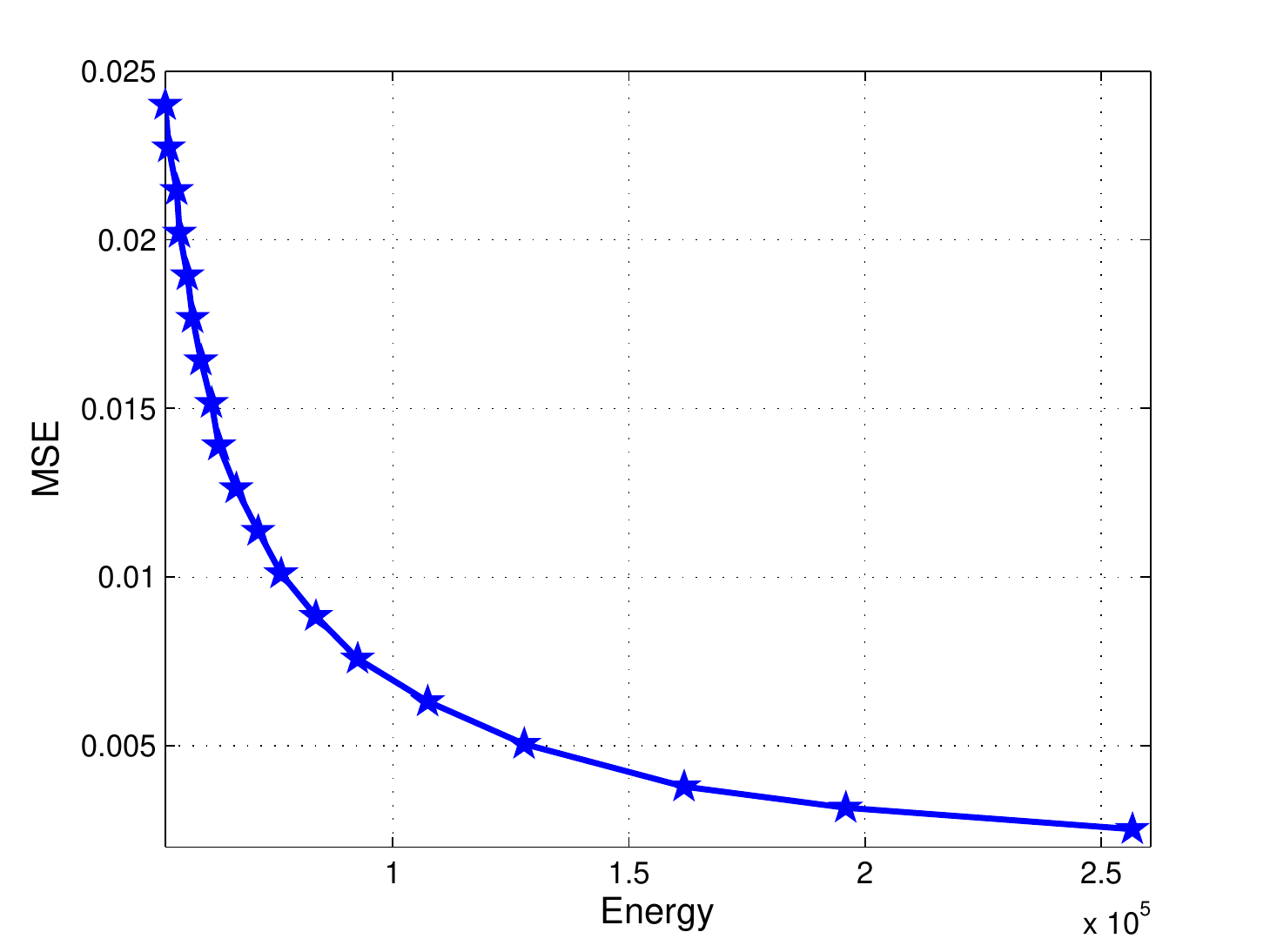}
	\caption{Pareto curve of energy consumption versus MSE levels obtained by \eqref{eq:myoptproblemconvexovertimeexplicit} with $\lambda = 10^3$ over $T = 10$ time instances for a fixed sensor network of $m = 100$ elements that estimate an unknown of size $n = 20$. The network topology we consider is shown in Fig. \ref{fig:network}.}
	\label{fig:figure7}
\end{figure}
Finally, we also show how the proposed algorithm can be applied to schedule a sensor network when absolute energy and communications costs are available. In terms of the sensing cost, we take it as
\begin{equation}
	s_i = O(\| \mathbf{a}_i^T \|_2^2),\ i = 1,\dots,m,
\end{equation}
meaning that the sensing cost is proportional to the quality of the measurement, its $\ell_2$ energy. The cost of the communications for the $i^\text{th}$ sensor is a fraction of the sensing cost
\begin{equation}
	c_{ii} = O(s_i),\ i = 1,\dots,m.
\end{equation}
We take the sensor network topology in Fig. \ref{fig:network} to which we attach a Gaussian random measurement matrix. We will consider that no reference energy levels are available, i.e., $\mathbf{e}_0 = \mathbf{0}_{m \times 1}$, and that the energy penalty is reflected by the cost function $g(\mathbf{e}) = \| \mathbf{e} \|_2^2$ in \eqref{eq:myoptproblemconvexovertimeexplicit}. In Fig. \ref{fig:figure7} we show the trade-off between the achievable MSE levels versus the energy consumption of the sensor network. To achieve the lowest levels of MSE we of course need (almost) all the sensors to be active (almost) all the time and therefore the energy consumption of the network is highest. Giving up some accuracy in the MSE has a positive impact on the energy consumption, especially at the limit of the best accuracy. Depending on the available energy supplies, Fig. \ref{fig:figure7} shows what levels estimation accuracy in terms of MSE are possible with the sensor network.

Regarding the running time, although it exhibits polynomial complexity, the proposed method is slower than some of the state-of-the-art methods from the literature, especially the ones based on greedy iterations. For example, for $m = 100$ and $n=20$, averaged over 100 realizations the MPME \cite{MPME2016} is computationally efficient with running times well below one second while the proposed method take about one minute to complete on a modern computing Intel i7$^{\textregistered}$ system. Therefore, the proposed method is now well suited for highly dynamical sensor network scheduling. As previously mentioned, the advantage of convex optimization based solutions is that they allow easy generalizations (like operating the sensor network over multiple time instances, without repetition of sensor selection) and allow the addition of extra constraints (like energy and communications).

\section{Conclusions}

In this paper we describe a new algorithm based on a convex optimization approach to deal with the sensor placement and scheduling problems. Our method is competitive against state-of-the-art sensor management methods while it also allows scheduling the network operations over time and with energy and communication costs and constraints. We are also able to show that when the sensor network measurements are given by a tight measurement matrix then we can expect the mean squared error of the estimation to decrease on average linearly with the number of active sensors. We also give a lower bound showing a potential quadratic decrease in the mean square error (in the best case scenario) with the number of active sensors. These theoretical insights into the sensor selection problem hold generally and are independent of the algorithmic approaches used. Furthermore, we show that sensor activation or scheduling with a tight measurement system is equivalent to sampling set selection for bandlimited graph signals and therefore the results presented in this paper are also relevant to the field of graph signal processing.

\appendices

\section{Proof of Result 2}

We use the determinant lemma $\det(\mathbf{A}^T \mathbf{A} + \mathbf{aa}^T) = \det(\mathbf{A}^T\mathbf{A}) (1 + \mathbf{a}^T (\mathbf{A}^T \mathbf{A})^{-1} \mathbf{a})$ to reach
\begin{equation*}
\begin{aligned}
\mu_{1} = & \frac{\lambda_1 (\lambda_1 + r)\dots (\lambda_1 + (n-1)r)}{\mu_2 \dots \mu_n} (1+\mathbf{a}^T (\mathbf{A}^T \mathbf{A})^{-1}\mathbf{a}) \\
\geq & \frac{\lambda_1 (\lambda_1 + r)\dots (\lambda_1 + (n-1)r)}{\lambda_1 \dots (\lambda_1 + (n-2)r)} (1+\mathbf{a}^T (\mathbf{A}^T \mathbf{A})^{-1}\mathbf{a}) \\
= & (\lambda_1 +(n-1)r) (1+\mathbf{a}^T (\mathbf{A}^T\mathbf{A})^{-1}\mathbf{a}).\\
%= & \frac{\lambda_1 \lambda_{\max}}{\lambda_1 + (n-1)r} (1+\mathbf{b}^T \mathbf{A}^{-1}\mathbf{b}).
\end{aligned}
\end{equation*}
The inequality holds when $\mu_i = \lambda_{i-1},\ i = 2,\dots,n$ -- the maximum values for $\mu_i$ according to \eqref{eq:interlacing}. For example, when $r=0$ all the eigenvalues of $\mathbf{A}^T\mathbf{A}$ are the same $\lambda_1$ and we have that $\mu_i = \lambda_1,\ i=2,\dots,n$ while $\mu_1 = (1+\mathbf{a}^T \mathbf{A}^{-1}\mathbf{a}) \lambda_1$.

\section{Proof of Result 3}

This qualitative result follows straight forward from the fact that the eigenvalues of $\mathbf{A}^T \mathbf{A}$ are all smaller or equal than the eigenvalues of $\mathbf{\tilde{A}}^T \mathbf{\tilde{A}} = \mathbf{A}^T \mathbf{A} + \mathbf{aa}^T$, by Result 1.

A quantitative analysis can also be made for the performance indicators. For the VCE we can use the determinant inversion lemma
\begin{equation*}
\begin{aligned}
\det(\mathbf{A}^T \mathbf{A} + \mathbf{aa}^T) = & \det(\mathbf{A}^T \mathbf{A})(1+ \mathbf{a}^T (\mathbf{A}^T \mathbf{A})^{-1} \mathbf{a}) \\
\geq & \det(\mathbf{A}^T \mathbf{A}), \forall\ \mathbf{a} \in \mathbb{R}^n.
\end{aligned}
\end{equation*}
For the MSE we can use the matrix inversion lemma
\begin{equation*}
\begin{aligned}
\text{tr}((\mathbf{A}^T \mathbf{A} + \mathbf{aa}^T )^{-1} ) = & \text{tr}( (\mathbf{A}^T \mathbf{A})^{-1} ) - \frac{\| (\mathbf{A}^T \mathbf{A})^{-1} \mathbf{a} \|_2^2 }{1 + \mathbf{a}^T (\mathbf{A}^T \mathbf{A})^{-1} \mathbf{a}} \\
\leq & \text{tr}( (\mathbf{A}^T \mathbf{A})^{-1} ), \forall \ \mathbf{a} \in \mathbb{R}^n.
\end{aligned}
\end{equation*}
In the case of the WCE we can bound the least singular value. Given a matrix $\mathbf{A} \in \mathbb{R}^{m \times n}$ and a row $\mathbf{a} \in \mathbb{R}^{n}$ then for extended matrix $\mathbf{\tilde{A}}^T = \begin{bmatrix} \mathbf{A}^T & \mathbf{a} \end{bmatrix} \in \mathbb{R}^{n \times (m+1)}$ we have
\begin{equation}
\! \! \left( \lambda_n^{-1/2}(\mathbf{A}^T\mathbf{A}) - \frac{ \| (\mathbf{\tilde{A}}^T \mathbf{\tilde{A}})^{-1} \mathbf{a} \|_2^2}{1-\mathbf{a}^T (\mathbf{\tilde{A}}^T \mathbf{\tilde{A}})^{-1} \mathbf{a}} \right)^{-2} \! \!  \geq \! \! \lambda_n(\mathbf{\tilde{A}}^T\mathbf{\tilde{A}}).
\label{eq:lambdaminbound}
\end{equation}

To show this we start by defining the smallest singular value
\begin{equation*}
\! \! \! \sigma^{-1}_{\min}(\mathbf{A}) \! = \! \sqrt{\lambda_n^{-1} ( \mathbf{A}^T\mathbf{A} )} \! = \! \| \mathbf{A}^{-1} \|_2 \! = \! \sqrt{\lambda_1 ( (\mathbf{A}^T\mathbf{A})^{-1} )}.
\end{equation*}
We use the fact that $\mathbf{A}^T\mathbf{A} = \mathbf{\tilde{A}}^T\mathbf{\tilde{A}} - \mathbf{aa}^T$ and use the Sherman-Morrison-Woodbury formula $(\mathbf{A}^T\mathbf{A})^{-1} = (\mathbf{\tilde{A}}^T\mathbf{\tilde{A}} - \mathbf{aa}^T)^{-1} = (\mathbf{\tilde{A}}^T\mathbf{\tilde{A}}  )^{-1} + \frac{( \mathbf{\tilde{A}}^T\mathbf{\tilde{A}})^{-1}  \mathbf{aa}^T  ( \mathbf{\tilde{A}}^T\mathbf{\tilde{A}})^{-1}}{1-\mathbf{a}^T (\mathbf{\tilde{A}}^T\mathbf{\tilde{A}})^{-1} \mathbf{a}} $ to reach $ \sigma_{\min}^{-1}(\mathbf{\tilde{A}}) \geq \sigma_{\min}^{-1}(\mathbf{A}) - \frac{ \| (\mathbf{\tilde{A}}^T \mathbf{\tilde{A}})^{-1} \mathbf{a} \|_2^2}{1-\mathbf{a}^T (\mathbf{\tilde{A}}^T \mathbf{\tilde{A}})^{-1} \mathbf{a}}$. Result \eqref{eq:lambdaminbound} follows directly from this last inequality.

In the special case of an invertible matrix $\mathbf{A} \in \mathbb{R}^{n \times n}$ and a row $\mathbf{a} \in \mathbb{R}^{n}$ then for extended matrix $\mathbf{\tilde{A}}^T = \begin{bmatrix} \mathbf{A}^T & \mathbf{a} \end{bmatrix} \in \mathbb{R}^{n \times (n+1)}$ we have
\begin{equation*}
\! \lambda_n(\mathbf{A}^T \mathbf{A}) \! \leq \! \lambda_n(\mathbf{\tilde{A}}^T \mathbf{\tilde{A}}) \! \leq \! \left(1+ \| \mathbf{a}^T \mathbf{A}^{-1}\|_2^2 \right) \lambda_n(\mathbf{A}^T \mathbf{A}).
\end{equation*}

To show this we start by developing
\begin{equation*}
\mathbf{\tilde{A}} = \begin{bmatrix} \mathbf{A} \\ \mathbf{a}^T \end{bmatrix} = \begin{bmatrix} \mathbf{A} \\ \mathbf{a}^T \mathbf{A}^{-1} \mathbf{A} \end{bmatrix} = \begin{bmatrix} \mathbf{I} \\ \mathbf{a}^T \mathbf{A}^{-1} \end{bmatrix} \mathbf{A}.
\end{equation*}
We know that $\sigma_{\min}(\mathbf{XY}) \leq \| \mathbf{X} \|_2 \sigma_{\min}(\mathbf{Y})$ by the Courant-Fischer-Weyl min-max principle. Coupled with the fact that the matrix $\begin{bmatrix}
\mathbf{I} & (\mathbf{a}^T \mathbf{A}^{-1})^T
\end{bmatrix}^T$ has all singular values 1 expect for the largest which is $\sqrt{1+\| \mathbf{a}^T \mathbf{A}^{-1}\|_2^2}$ we reach the result. The final inequality follows from
\begin{equation*}
1 + \| \mathbf{a}^T \mathbf{A}^{-1} \|_2^2 \leq 1 + \sigma_{\min}^{-2} (\mathbf{A}) \| \mathbf{a}\|_2^2.
\end{equation*}
The equality holds when we choose $\mathbf{a}$ to be any multiple of the right singular vector associated with $\sigma_{\min}(\mathbf{A})$. Overall the inequalities become $\sigma_{\min}(\mathbf{A}) \leq \sigma_{\min}(\mathbf{\tilde{A}}) \leq  \left( \sqrt{1+ \| \mathbf{a}^T \mathbf{A}^{-1}\|_2^2} \right) \sigma_{\min}(\mathbf{A})
\leq \sqrt{\sigma_{\min}^2(\mathbf{A})+\|\mathbf{a} \|_2^2}$.
Together with Result 1 we ultimately have that
\begin{equation*}
\sigma_{\min}(\mathbf{\tilde{A}}) \leq  \min \left\{ \sqrt{\sigma_{\min}^2(\mathbf{A})+\|\mathbf{a} \|_2^2},\ \sigma_{n-1} \right\}.
\end{equation*}

\section{Proof of Result 4}

We will use results presented in \cite{Ramanujan2013} to prove theorems about the expected characteristic polynomials of matrices that are changed by rank 1 updates. Consider first that for the $\alpha-$tight frame $\mathbf{A}^T = \begin{bmatrix} \mathbf{a}_1 & \mathbf{a}_2 & \dots & \mathbf{a}_m \end{bmatrix} \in \mathbb{R}^{n \times m}$ and given any vector $\mathbf{u}$ such that $\| \mathbf{u} \|_2 = 1$ we have that
\begin{equation}
%\begin{aligned}
\mathbb{E}[ (\mathbf{u}^T \mathbf{a}_\text{avg})^2 ] = \frac{1}{m}\sum_{i=1}^m (\mathbf{u}^T \mathbf{a}_i)^2
%= \frac{1}{m} \mathbf{u}^T \left( \sum_{i=1}^m \mathbf{a}_i \mathbf{a}_i^T \right) \mathbf{u}
= \frac{\alpha \| \mathbf{u} \|_2^2}{m} = \frac{\alpha}{m},
%\end{aligned}
\label{eq:Eofaavg}
\end{equation}
where we have defined the average frame vector $\mathbf{a}_\text{avg} = \frac{1}{\sqrt{m}} \sum_{i=1}^m \mathbf{a}_i$.

If we denote by $p_{\mathbf{A}_1^T \mathbf{A}_1}(x)$ the characteristic polynomial of $\mathbf{A}_1^T \mathbf{A}_1$ then the characteristic polynomial of $\mathbf{A}_1^T\mathbf{A}_1 + \mathbf{aa}^T$ is
\begin{equation*}
\begin{aligned}
p_{\mathbf{A}_1^T\mathbf{A}_1 + \mathbf{aa}^T}&(x) =  \det(x\mathbf{I} - (\mathbf{A}_1^T\mathbf{A}_1 + \mathbf{aa}^T)) \\
= & \det((x\mathbf{I} - \mathbf{A}_1^T\mathbf{A}_1) - \mathbf{aa}^T) \\
= & \det(x\mathbf{I} - \mathbf{A}_1^T\mathbf{A}_1)(1 - \mathbf{a}^T(x\mathbf{I} - \mathbf{A}_1^T\mathbf{A}_1)^{-1} \mathbf{a}) \\
= & p_{\mathbf{A}_1^T \mathbf{A}_1}(x) \left( 1 - \sum_{i=1}^n \frac{(\mathbf{u}_i^T \mathbf{a})^2}{x - \lambda_i(\mathbf{A}_1^T\mathbf{A}_1)} \right),
\end{aligned}
\end{equation*}
where $\lambda_i(\mathbf{A}_1^T\mathbf{A}_1)$ are the eigenvalues of $\mathbf{A}_1^T\mathbf{A}_1$ corresponding to eigenvectors $\mathbf{u}_j$. Denoting by $\mathcal{K}$ the set of indices already selected, if we choose any $\mathbf{a} \in \{ \mathbf{a}_i \}_{i=1, i \notin \mathcal{K}}^m$ and use \eqref{eq:Eofaavg} we can show that
\begin{equation*}
\begin{aligned}
\mathbb{E}[p_{\mathbf{A}_1^T\mathbf{A}_1 + \mathbf{aa}^T}(x)] = & \mathbb{E}[p_{\mathbf{A}_1^T\mathbf{A}_1}(x)] \left( 1 -  \sum_{i=1}^n \frac{\alpha m^{-1}}{x - \lambda_i(\mathbf{A}_1^T\mathbf{A}_1)} \right) \\
= & \mathbb{E}[p_{\mathbf{A}_1^T \mathbf{A}_1}(x)] - \frac{\alpha}{m} \mathbb{E}[p_{\mathbf{A}_1^T \mathbf{A}_1}(x)]^\prime.
\end{aligned}
\end{equation*}

Starting from an empty (all zeros) matrix $\mathbf{A}_1^T\mathbf{A}_1$, i.e., $\mathcal{K} = \emptyset$, that has the characteristic polynomial $p_{\mathbf{A}_1^T\mathbf{A}_1}^{(0)}(x) =p_{\mathbf{0}_{n \times n}}(x) = x^n$ after adding $k$ rank 1 updates of the type $\mathbf{a} \mathbf{a}^T$ leads to the matrix $\mathbf{A}_1^T\mathbf{A}_1$ with the expected characteristic polynomial is
\begin{equation*}
\begin{aligned}
\mathbb{E}[p_{\mathbf{A}_1^T \mathbf{A}_1}^{(k)}(x)] = & \mathbb{E}[p_{\mathbf{A}_1^T \mathbf{A}_1}^{(k-1)}(x)] - \frac{\alpha}{m} \mathbb{E}[p^{ (k-1)}_{\mathbf{A}_1^T \mathbf{A}_1}(x)]^\prime\\
= & a_n^{(k)} x^n + \dots + a_1^{(k)} x + a_0^{(k)}.
\end{aligned}
%\label{eq:expectedpolynomial}
\end{equation*}
The results in \eqref{eq:averageperformance} follow from Vieta's formulas that relate roots of polynomials to their coefficients. The expected value of the $\text{VCE}(\mathbf{A}_1)$ follows directly as the constant coefficient of the expected characteristic polynomial since it is the product of the roots while for the $\text{MSE}(\mathbf{A}_1)$ we have
\begin{equation*}
\sum_{i=1}^n \frac{1}{\lambda_i(\mathbf{A}_1^T \mathbf{A}_1)} = \frac{\prod_{i \neq 1} \lambda_i(\mathbf{A}_1^T \mathbf{A}_1) + \dots + \prod_{i \neq n} \lambda_i(\mathbf{A}_1^T \mathbf{A}_1)}{\prod_{i=1}^n \lambda_i(\mathbf{A}_1^T \mathbf{A}_1)}.
\end{equation*}
From this it follows that
\begin{equation*}
\begin{aligned}
&\mathbb{E}[\text{MSE}(\mathbf{A}_1)] = -\frac{a_1^{(k)}}{a_0^{(k)}},\ \mathbb{E}[\text{WCE}(\mathbf{A}_1)] \geq -\frac{a_1^{(k)}}{a_0^{(k)} n}, \\ &\quad \quad \quad \quad \quad \mathbb{E}[\text{VCE}(\mathbf{A}_1)] = \log ( a_0^{(k)} ).
\end{aligned}
%\label{eq:averageperformance2}
\end{equation*}
Notice that for $k = n$ the expected characteristic polynomial is an associated Laguerre polynomial $n! L_n(x)$ \cite{Laguerre2006} with coefficients
\begin{equation*}
a_i^{(n)} = \frac{(-1)^i n!}{i!} \left( \frac{\alpha}{m} \right)^{n-i} {n \choose i}, \ i=0,\dots,n.
\end{equation*}
Then for $k \geq n$ we have the coefficients of interest
\begin{equation}
a_0^{(k)} \! \! = \! \! n! \left( \frac{\alpha}{m} \right)^n {k \choose k-n},
a_1^{(k)} \! \! = \! \! -n! \left( \frac{\alpha}{m} \right)^{n-1} {k \choose k-n+1}.
\label{eq:inbetween}
\end{equation}
Since we are dealing with positive semidefinite matrices the expected values have to be positive and therefore once we have written the values for $a_0^{(k)}$ and $a_1^{(k)}$ results in \eqref{eq:averageperformance} follow immediately from \eqref{eq:inbetween}.

\bibliographystyle{IEEEtran}
\bibliography{refs}

\end{document}